%% file: STEP1.tex
\documentclass[useAMS,usenatbib,times]{mn2e}
\usepackage{epsfig,times,amsmath, amsfonts, amssymb}

\title[STEP: Weak lensing analysis of simulated ground-based observations]{The
  Shear TEsting Programme 1: Weak lensing analysis of simulated ground-based
  observations} 
\author[Heymans et al.]{Catherine Heymans$^{1}$\thanks{heymans@physics.ubc.ca},
Ludovic Van Waerbeke$^{2}$, David Bacon$^{3}$, Joel Berge$^{4}$, 
\newauthor Gary Bernstein$^{5}$, Emmanuel
Bertin$^{6}$, Sarah Bridle$^{7}$, Michael L. Brown$^{3}$, 
\newauthor Douglas Clowe$^{8}$, H{\aa}kon Dahle$^{9}$, 
Thomas Erben$^{10}$, 
Meghan Gray$^{11}$, Marco Hetterscheidt$^{10}$, \newauthor 
Henk Hoekstra$^{12}$, Patrick Hudelot$^{13}$, Mike Jarvis$^{5}$, 
Konrad Kuijken$^{14}$,
Vera Margoniner$^{15}$,\newauthor 
Richard Massey$^{16}$, 
Yannick Mellier$^{6,17}$,
Reiko Nakajima$^{5}$, Alexandre Refregier$^{4}$, \newauthor 
Jason Rhodes$^{18}$, Tim Schrabback$^{10}$ \&
 David Wittman$^{15}$.\\
$^1$Max-Planck-Institut f\"{u}r Astronomie, K\"{o}nigstuhl, D-69117,
Heidelberg, Germany.\\
$^2$University of British Columbia, 6224 Agricultural Rd., Vancouver, BC, V6T
  1Z1, Canada.\\  
$^3$Institute for Astronomy, University of Edinburgh, Royal Observatory,
Blackford Hill, Edinburgh, EH9 3HJ, UK. \\
$^{4}$ Service d'Astrophysique, CEA Saclay, F-91191 Gif sur Yvette, France.\\
$^5$Department of Physics and Astronomy, University of Pennsylvania,
  Philadelphia, PA 19104, USA.\\ 
$^6$Institut d'Astrophysique de Paris, UMR7095 CNRS,
   Universit\'e Pierre \& Marie Curie, 98 bis boulevard Arago, 
75014 Paris, France.\\
$^7$Department of Physics and Astronomy, University College London, Gower
  Street, London, WC1E 6BT, UK.\\
$^8$Steward Observatory, University of Arizona, 933 N. Cherry Ave., Tuscon,
    AZ 85721, USA.\\
$^9$Institute of Theoretical Astrophysics, University of Oslo, P.O. Box 1029,
  Blindern, N-0315 Oslo, Norway.\\ 
$^{10}$Institut f\"ur Astrophysik und Extraterrestrische Forschung, Universit\"at
  Bonn, Auf dem H\"ugel 71, 53121 Bonn, Germany.\\
$^{11}$School of Physics and Astronomy, University of Nottingham, Nottingham,
  NG7 2RD, UK.\\ 
$^{12}$University of Victoria, Elliott Building, 3800 Finnerty Rd, Victoria,
  BC, V8P 5C2, Canada.\\
$^{13}$Observatoire Midi-Pyr\'en\'ees, UMR5572, 14 Avenue Edouard Belin, 31000
  Toulouse, France.\\
$^{14}$Leiden Observatory, P.O. Box 9513, NL-2300 RA, Leiden, The
  Netherlands. \\  
$^{15}$Department of Physics, University of California at Davis, One Shields
Avenue, Davis, CA 95616, USA.\\
$^{16}$California Institute of Technology, Pasadena, CA 91125, USA.\\
$^{17}$Observatoire de Paris. LERMA. 61, avenue de l'Observatoire. 75014 Paris,
France.\\
$^{18}$ Jet Propulsion Laboratory, 4800 Oak Grove Drive, Pasadena, CA
  91109, USA.\\ }

\newcommand{\be}{\begin{equation}}  \newcommand{\ee}{\end{equation}}
\newcommand{\bes}{\begin{equation*}}  \newcommand{\ees}{\end{equation*}}
  \newcommand{\ba}{\begin{eqnarray}}
\newcommand{\ea}{\end{eqnarray}}  
\newcommand{\nn}{\nonumber\\}  

\newcommand{\rag}{\rangle}
\newcommand{\lag}{\langle}
\newcommand{\bm}[1]{\mbox{\boldmath{$#1$}}}
\def\gs{\mathrel{\raise1.16pt\hbox{$>$}\kern-7.0pt 
\lower3.06pt\hbox{{$\scriptstyle \sim$}}}}         
\def\ls{\mathrel{\raise1.16pt\hbox{$<$}\kern-7.0pt 
\lower3.06pt\hbox{{$\scriptstyle \sim$}}}}         

\begin{document}

\pagerange{\pageref{firstpage}--\pageref{lastpage}} \pubyear{2005}
\maketitle
\label{firstpage}

\begin{abstract}

The Shear TEsting Programme, STEP, is a collaborative project
to improve the accuracy and reliability of 
all weak lensing measurements in preparation for the next 
generation of wide-field surveys.  In this first STEP paper
we present the results of a blind analysis of simulated 
ground-based observations of relatively simple galaxy morphologies.
The most successful methods are shown to achieve percent
level accuracy. From the cosmic shear pipelines that have been
used to constrain cosmology, we find weak lensing shear measured to an accuracy
that is within the 
statistical errors of current weak lensing analyses, with shear
measurements accurate to
better than 7\%.  The dominant source of measurement error is shown to arise
from calibration uncertainties where the measured shear is
over or under-estimated by a constant multiplicative factor.  This is of
concern as calibration errors cannot be detected
through standard diagnostic tests.  The measured calibration
errors appear to result from 
stellar contamination, false object detection, 
the shear measurement method itself,
selection bias and/or the use of biased weights.
Additive systematics (false detections of shear) 
resulting from residual point-spread function anisotropy 
are, in most cases, 
reduced to below an equivalent shear of 0.001, 
an order of magnitude
below cosmic shear distortions on the scales probed by current surveys.

Our results provide a snapshot view of the accuracy of 
current ground-based weak lensing methods and a benchmark upon which we can
improve.  To this end 
we provide descriptions of each method tested and include details
of the eight different implementations of the commonly used  \citet{KSB}
method (KSB+) to aid the improvement of future KSB+ analyses.  

\end{abstract}


\section{Introduction}

Gravitational lensing provides an unbiased way to study the 
distribution of matter in the Universe.
Derived from the physics of gravity, where gravitational light
deflection is dependent solely on the distribution of matter,
weak gravitational lens theory describes a unique way to directly
probe dark matter on large scales \citep[see the extensive review
  by][]{Bible}.   This tool has many astronomical applications; 
the detection of weak shear around galaxy clusters
yields an estimate of the total cluster mass 
\citep[see for example][]{WittMargon,Margoniner}
and enables a full mass reconstruction of low redshift clusters \citep[see for
example][]{Clowe04, meg02, DahleK2K};  the average weak tangential 
shear of distant galaxies around nearby galaxies constrains the
ensemble average properties of dark matter halos \citep[see for
example][] {HoekstraGG,SheldonGG}; 
the weak lensing of background galaxies by foreground large-scale structure 
directly probes the evolution of the 
non-linear matter power spectrum, hence 
providing a signal that can constrain cosmological parameters \citep[see
  review by][]{VwBrev}.  This last application has the
great promise of being able to tightly constrain the properties of dark energy
with the next generation of wide-field multi-colour surveys
\citep{JainTay,BernsteinJain,Benabed,Heavens03,RefSNAP03}.

Technically, weak lensing is rather challenging to detect.  It
requires the measurement of the weak 
distortion that lensing induces in the shapes of observed 
galaxy images.  These images have
been convolved with the point spread
function (PSF) distortion of the atmosphere, telescope and camera. 
The accuracy of any analysis therefore depends critically 
on the correction for instrumental distortions and atmospheric seeing. 
Weak lensing by large-scale structure induces percent level
correlations in the observed ellipticities of galaxies, termed `cosmic shear'.
This cosmological application of weak lensing theory 
is therefore the most demanding technically,
owing to the fact that for any weak lensing survey, the 
instrumental distortions are an order of
magnitude larger than the underlying cosmic shear distortion that we
wish to detect.  We therefore focus on the demands of this particular 
application even
though our findings will be beneficial to all weak lensing studies.

The unique qualities of weak lensing as a dark matter and dark energy probe
demand that all technical challenges are met and overcome, and
this desire has lead to the development of some of the most innovative
methods in astronomy.  The first pioneering weak lensing measurement methods by
\citet{TysonWenkValdes}, \citet{BonMel} and \citet{KSB} (KSB) 
have improved \citep{LK97,HFKS98} (KSB+) and
diversified \citep{RRG00,K00,im2shape,Bernstein,shapelets,Masseyshapelets}.
Novel methods to model the spatial and temporal variation of the PSF 
have also been designed to improve the success of 
the PSF correction \citep{Hoekstra03,JarvisPSF}.
In addition, diagnostic techniques have been developed and implemented to
provide indicators for the presence of residual 
systematic non-lensing distortions \citep{BMRE,CNPT02,SchvWM02,MLB02}.  

Rapid technical development has mirrored the growth in observational
efforts with the cosmic shear 
analysis of several wide-field optical surveys yielding joint constraints on
the matter density parameter $\Omega_m$ and
the amplitude of the matter power spectrum $\sigma_8$
\citep{Maoli,RRG01,vWb01,HYG02,BMRE,RRG02,Jarvis,MLB02,Hamana,Massey,RhodesSTIS,vWb04,HymzGEMS,Jarvis05,CFHTLSwide,CFHTLSdeep} and also constraints on the dark energy equation of state parameter $w$ \citep{Jarvis05,CFHTLSwide,CFHTLSdeep}.
The results from these efforts are found to be in broad agreement and are
fast becoming more credible
with the most recent publications presenting the results from 
several different diagnostic tests to determine the levels of systematic
error.  Table~\ref{tab:sig8} lists the most recent cosmic shear results from
different authors or surveys, the two-point statistics used in the
cosmological parameter analysis and
the statistics used to determine levels of systematic errors through an
E/B mode decomposition \citep{CNPT02}.  
See \citet{SchvWM02} and \citet{MLB02}
for details about each two-point statistic and their E/B mode decomposition
and \citet{Massey}, \citet{vWb04} and \citet{HymzGEMS} for different 
discussions on which statistics are best to use.
For such a young field of observational research, the
$\sim 2\sigma$ agreement between the results, shown in Table~\ref{tab:sig8}, 
is rather impressive.  
The differences between the results are, however, 
often cited as a reason for caution
over the use of cosmic shear as a cosmological probe.  
For this reason the Shear TEsting
Programme\footnote{www.physics.ubc.ca/$\sim$heymans/STEP.html} (STEP) 
was launched in order to improve the
accuracy and reliability of all future 
weak lensing measurements through the rigorous
testing of shear measurement pipelines, the exchange of data and the sharing of
technical and theoretical knowledge within the weak lensing community. 

The current differences seen in cosmic shear cosmological 
parameter estimates could result from a
number of sources; inaccurate source redshift distributions that 
are required to
interpret the cosmic shear signal; sampling variance; systematic errors from
residual instrumental distortions; calibration biases in the shear measurement
method.  Contamination to cosmic shear analyses 
from the intrinsic galaxy alignment of nearby galaxies
is currently thought to be a weak effect that is measured and
mitigated in \citet{HBH04} \citep[also see][and references
  therein]{KingSch03,HH03,KingSch02}. 
With the next generation of wide-field multi-colour surveys
many of these problems can swiftly be resolved as the multi-colour
photometric redshifts will provide a good
estimate of the redshift distribution \citep[see for
  example][]{MLB02}
and the wide areas will minimise sampling
variance.  In addition, all new instrumentation
has been optimised to reduce the severity of instrumental distortions
improving the accuracy of future PSF corrections.
Implementing diagnostic statistics that decompose cosmic shear signals into
their lensing E-modes and non-lensing B-modes
\citep{CNPT02,SchvWM02,MLB02} immediately alerts us to the
presence of systematic error within our data set.  B-mode
systematics can then be reduced through the modification of PSF models
\citep{vWb04,JarvisPSF} or merely the
selection of angular scales above or below which the systematics are removed. 
Calibration bias is therefore perhaps of greatest concern as,
in contrast to additive PSF errors, it can only be directly 
detected through the cosmic
shear analysis of image simulations, although see 
the discussion on self-calibration in \citet{Huterer} and
\citet{HirataMandelbaum} and \citet{MandelbaumHirata} for 
model-dependent estimates of shear calibration errors in the Sloan Digital
Sky Survey.
With the statistics currently used to
place constraints on cosmological parameters, a shear 
calibration error contributes directly to an error in $\sigma_8$.  
The recent development of statistics which are fairly 
insensitive to shear calibration errors \citep{JainTay,Bernstein05}
are certainly one 
solution to this potential problem.  Also see \citet{Ishak}, where shear 
calibration uncertainties are marginalised over in the cosmological parameter
estimation.

\begin{table*}
\begin{center}
\begin{tabular}{l|l|c|c|c|c|c}
\bf{Survey Analysis} & \bf{Pipeline Description}& $\bm{\sigma_8}$ &
\bf{Statistic} & {\bf E/B decomposition} & {\bf Area}
  ($\bm{{\rm deg}^2}$) & $\bm{z_m}$ \\\hline\hline
\citet{HYG02} & \citet{HFKS98} & $0.86 ^{+0.09}_{-0.13}$ &
$\lag {\rm M_{\rm ap}}^2 \rag$ & $\lag {\rm M_{\rm ap}}^2 \rag$\,
                               $\lag {\rm M}_\perp^2 \rag$ 
& $53.0$ & $0.54 - 0.66$\\\hline
\citet{RRG02} & \citet{RRG00} & $0.94 \pm 0.24$ &
$\lag \gamma^2 \rag$ &
$\lag {\rm M_{\rm ap}}^2 \rag$\, $\lag {\rm M}_\perp^2 \rag$ 
& $0.36$ (s) & $0.9 \pm 0.1$\\\hline
\citet{MLB02} & \citet{BRE} & $0.72 \pm 0.09$  &
$\xi_{\pm}$\, $P^{\kappa\kappa}$ & $P^{\kappa\kappa}$\,$
  P^{\kappa\beta}$\, $P^{\beta\beta}$ 
& $1.25$ & $0.85 \pm 0.05$\\\hline
\citet{Hamana} & \citet{Hamana} & $0.78 ^{+0.55}_{-0.25}$ &
$\lag {\rm M_{\rm ap}}^2 \rag$ & $\lag {\rm M_{\rm ap}}^2 \rag$\, 
                               $\lag {\rm M}_\perp^2 \rag$ 
& $ 2.1$ & $0.6 - 1.4$\\\hline
\citet{RhodesSTIS} & \citet{RRG00} & $1.02 \pm 0.16 $ &
$\lag \gamma^2 \rag$ & none 
& $0.25$ (s) & $1.0 \pm 0.1$\\ \hline 
\citet{vWb04} & \citet{vWb00} & $0.83 \pm 0.07$ &
$\lag {\rm M_{\rm ap}}^2 \rag$\, $\xi^{\rm E}$ & $\lag {\rm M_{\rm ap}}^2 \rag$\, 
                               $\lag {\rm M}_\perp^2 \rag$\, $\xi^{\rm
  E}$\,$\xi^{\rm B}$
& $8.5$ & $0.8 - 1.0$\\\hline 
\citet{Jarvis05} & \citet{Bernstein} & $0.72^{+0.17}_{-0.14}$ &
$\lag \gamma^2 \rag$\, $\lag {\rm M_{\rm ap}}^2 \rag$&
$\lag {\rm M_{\rm ap}}^2 \rag$\, $\lag {\rm M}_\perp^2 \rag$
& $75.0$ & $0.6 \pm 0.1$\\\hline
\citet{Massey}& \citet{BRE} & $1.02 \pm 0.15$ & 
$\xi_{\pm}$ & $\xi^{\rm E}$\,$\xi^{\rm B}$
& $4.5$ & $0.8 \pm 0.08$\\\hline
\citet{HymzGEMS} & \citet{HymzGEMS} & $0.68 \pm 0.13$ &
$\xi_{\pm}$, $P^{\kappa\kappa}$ &  $\xi^{\rm
  E}$\,$\xi^{\rm B}$\,$P^{\kappa\kappa}$\,$
  P^{\kappa\beta}$\, $P^{\beta\beta}$ & $0.22$ (s) & $1.0 \pm 0.1$ \\\hline\hline
\end{tabular}
\end{center}
\caption{The most recent cosmological parameter constraints on the amplitude
  of the matter power spectrum $\sigma_8$ 
from each author or survey, for a matter density parameter 
$\Omega_m = 0.3$.  Quoted errors on $\sigma_8$ are $1\sigma$ (68\% confidence)
  except in the case of \citet{Jarvis05} where the errors given are $2\sigma$
  (95\% confidence).  Several different statistics have been used to 
constrain $\sigma_8$, as detailed, where  $\lag {\rm
  M_{\rm ap}}^2 \rag$ is the mass aperture statistic, $\lag \gamma^2 \rag$ is 
the top-hat shear variance, $\xi_{\pm}$ are the
  shear correlation functions and $P^{\kappa\kappa}$ is the shear power
  spectrum.  The statistics used to determine the level of non-lensing B-modes
  in each result are also listed where $\lag {\rm
  M_{\perp}}^2 \rag$ is the B-mode mass aperture statistic, $\xi^{\rm
  E}$ and $\xi^{\rm B}$ are E and B mode correlators, 
$P^{\beta\beta}$ is the B-mode shear power spectrum, and $P^{\kappa\beta}$ is
  the E/B cross power spectrum.  See \citet{SchvWM02} and \citet{MLB02}
for details about each two-point statistic and their E/B mode decomposition. 
The shear measurement
pipeline that has been used for each result is listed for reference, along
  with the area of the survey and the median redshift estimate of the survey 
$z_m$.  Space-based surveys are denoted with an (s) in the area column.}
\label{tab:sig8}
\end{table*}
 
\citet{Baconsims}, \citet{erben} and \citet{HYGBHI}
presented the first detailed cosmic shear 
analyses of artificial 
image simulations using the KSB+ method. \citet{Baconsims} found
that the KSB+ method was reliable to $\sim 5\%$ 
provided a calibration factor of $0.85$ was included in the analysis
to increase the KSB+ shear estimator.  The calibration factor has since been
included in the work of \citet{BMRE}, \citet{MLB02} and \citet{Massey} 
who implement the 
KSB+ pipeline tested in \citet{Baconsims}.  \citet{erben} found that depending
on the PSF type tested and the chosen 
implementation of the KSB+ formula, described in
section~\ref{sec:KSBmeth}, the KSB+ method 
was reliable to $\pm 10-15\%$ and did not require a calibration correction.
The artificial images tested by \citet{HYGBHI} included cosmic shear derived 
from ray-tracing simulations.  They found that the input lensing signal could
be recovered to better than $10\%$ of the input value.
The difference between these three conclusions is important. 
All papers adopted the same KSB+ method, but subtle differences in their
implementation resulted in the need for a calibration correction
in one case but not in the others.  It is therefore not
sufficient to cite these papers to support the KSB+ method as every
individuals' KSB+ pipeline implementation may differ slightly, 
introducing a discrepancy between the results.  

For the cosmic shear, galaxy-galaxy lensing 
and cluster mass determinations published to
date, $\le10\%$ errors are at worst comparable to the
statistical errors and are not dominant.  Much larger surveys now underway
will,
however, reduce statistical errors on various shear measurements to
the $\sim2\%$ level, requiring shear measurement accurate to $\sim1\%$.
In the next decade, deep weak-lensing surveys of thousands of square
degrees will require shear measurements accurate to $\sim 0.1\%$.
The technical challenges associated with measuring weak lensing shear must 
therefore be addressed and solved in a relatively short period of time.

Whilst KSB+ is currently 
the most widely used weak lensing method, promising alternative
methods have been developed 
[\citealt{RRG00} (RRG); \citealt{K00} (K2K); \citealt{Smith_thesis} ({\it
    ellipto}); \citealt{im2shape} (Im2shape);
\citealt{Bernstein} (BJ02); \citealt{shapelets1} (shapelets); 
\citealt{Masseyshapelets} (polar shapelets)] and
implemented in cosmic shear analyses [see for example \citealt{RhodesSTIS}
(RRG); \citealt{Witt01} ({\it ellipto}); \citealt{Jarvis} and
  \citealt{Jarvis05}  (BJ02); \citealt{Chang} 
  (shapelets)], and cluster lensing studies [see for example
  \citealt{im2shapeimp} (Im2shape); \citealt{DahleK2K} (K2K); 
\citealt{Margoniner} ({\it ellipto})].  Thorough testing of these 
newer techniques 
is however somewhat lacking in the literature, although see
\citet{shapelets} and \citet{polshapetest} for 
tests of the shapelets method.  

In this paper we present the first of the STEP initiatives; the
blind\footnote{CH, LV and KK knew the input shear of the simulations.}
analysis of 
sheared image simulations with a variety of 
weak lensing measurement pipelines used by
each author in their previously published work.  Authors and methods
are listed in Table~\ref{tab:methods}.  Modifications to pipelines used in
published work have not been allowed in light of
the results and we thus present our results openly to provide
the reader with a snapshot view of how accurately we can currently measure weak
lensing shear from galaxies with relatively simple morphologies.  
This paper will thus provide a
benchmark upon which we can improve in future STEP initiatives.  
Note that some of the methods evaluated in this paper
are experimental and/or in early stages
of development, notably the methods of \citet{KK06}, the
deconvolution fitting method of Nakajima (2005 in preparation), and the Dahle
implementation of K2K.   
The results from these particular methods should therefore not be taken as 
a judgment on their ultimate potential.

\begin{table}
\begin{center}
\begin{tabular}{lll}
\bf{Author} & \bf{Key} & \bf{Method}  \\\hline\hline
Bridle \& Hudelot & SB & Im2shape \citep{im2shape} \\\hline
Brown & MB& KSB+ [\citet{BRE} pipeline]\\\hline
Clowe &  C1 \& C2 & KSB+  \\\hline
Dahle & HD & K2K \citep{K00}\\\hline
Hetterscheidt & MH & KSB+ [\citet{erben} pipeline] \\\hline
Heymans & CH & KSB+ \\\hline
Hoekstra & HH & KSB+ \\\hline
Jarvis & MJ & 
\citet{Bernstein} \\
 & & Rounding kernel method \\ \hline
Kuijken & KK & Shapelets to $12^{th}$ order\\
               &    & \citet{KK06} \\\hline
Margoniner & VM & \citet{Witt01} \\\hline
Nakajima & RN & \citet{Bernstein} \\
 & & Deconvolution fitting method \\ \hline
Schrabback & TS & KSB+\\ 
 & & [\citet{erben} + modifications]\\\hline
Van Waerbeke & LV & KSB+ \\\hline\hline
\end{tabular}
\end{center}
\caption{Table of authors and methods.  The key identifies the authors
  in all future plots and Tables.} 
\label{tab:methods}
\end{table}

This paper is organised as follows. In
Section~\ref{sec:methods}  we review the different shear measurement methods
used by each author and describe the simulated data set in
Section~\ref{sec:data}.  We compare each authors' measured shear with the input
simulation shear in Section~\ref{sec:analysis} investigating forms of
calibration bias, selection bias and weight bias.  Note that our discussion on
the issue of source selection bias is indeed relevant for many different
types of survey analysis, not only the lensing applications detailed here. 
We discuss our findings in Section~\ref{sec:diss} and
conclude in Section~\ref{sec:conc}.

\section{Methods}
\label{sec:methods}
In the weak lensing limit the ellipticity of a galaxy is an unbiased estimate
of the gravitational shear.  For a perfect ellipse with axial ratio $\beta$ at
position angle $\theta$, measured counter-clockwise from the $x$ axis, we can
define the following ellipticity parameters \citep{BonMel}
\be
\left(
\begin{array}{c}
e_1 \\
e_2
\end{array}
\left)
= \frac{1-\beta}{1+\beta}
\right(
\begin{array}{c}
\cos 2\theta \\
\sin 2 \theta
\end{array}
\right),
\label{eqn:elliparam}
\ee
and the complex ellipticity $e = e_1 + i e_2$.  In the case of weak shear
$|\gamma| \ll 1$, the shear $\gamma = \gamma_1 + i\gamma_2$ is directly 
related to the average galaxy
ellipticity, $\gamma \approx \lag e \rag $.  
In this section we briefly review the different measurement methods used
in this STEP analysis to estimate galaxy ellipticity in the presence of
instrumental and atmospheric distortion and hence obtain an estimate of the 
gravitational shear $\gamma$.  Common to all methods is the initial source
detection stage, typically performed using the {\it 
  SExtractor} \citep{SExt} software.  The peak finding tool {\it
    hfindpeaks} from the {\it imcat}\footnote{ {\it 
    www.ifa.hawaii.edu/$\sim$kaiser/imcat/}} software is used as an alternative
in some KSB+ methods, listed in Appendix Table~\ref{tab:KSB_details}.  In
order to characterise the PSF, stars are selected in all cases
from a magnitude-size plot. 

\subsection{KSB+ Method}
\label{sec:KSBmeth}
\citet{KSB}, \citet{LK97} and \citet{HFKS98} (KSB+) prescribe a method to
invert the effects of 
the PSF smearing and shearing, recovering a shear
estimator uncontaminated by the systematic distortion of the PSF. 

Objects are parameterised according to their weighted quadrupole moments
\be
Q_{ij} = \frac{\int \, d^2\theta \, W(\bm{\theta}) \,
  I(\bm{\theta}) \, \theta_i
  \theta_j} {\int d^2\theta \, W(\bm{\theta}) \,I(\bm{\theta}) },
\label{eqn:quadmom}
\ee
where $I$ is the surface brightness of the object, $\theta$ is the angular
distance from the object centre and $W$ is a Gaussian weight function of scale
length $r_g$, where $r_g$ is some measurement of galaxy size.  For a
perfect ellipse, the weighted quadrupole moments are related to the
weighted ellipticity parameters\footnote{The KSB+ definition of
galaxy ellipticity differs from equation~\ref{eqn:elliparam}.
If the weight function $W(\bm{\theta}) = 1$ in equation~\ref{eqn:quadmom},
the KSB+ ellipticity $|\varepsilon| = (1-\beta^2)/(1+\beta^2)$, where $\beta$
is the axial ratio \citep[see][]{Bible}.} $\varepsilon_\alpha$ by
\be
\left(
\begin{array}{c}
\varepsilon_1 \\
\varepsilon_2
\end{array}
\left)
= \frac{1}{Q_{11} + Q_{22}}
\right(
\begin{array}{c}
Q_{11} - Q_{22} \\
2Q_{12}
\end{array}
\right).
\label{eqn:ellipquad}
\ee
\citet{KSB} show that if the PSF
distortion can be described as a small but highly anisotropic
distortion convolved with a large circularly symmetric seeing disk,
then the ellipticity of a PSF corrected galaxy is given by
\be
\varepsilon^{\rm cor}_{\alpha} = \varepsilon^{\rm obs}_{\alpha} - P^{\rm
  sm}_{\alpha\beta}p_{\beta}, 
\label{eqn:ecor}
\ee
where $p$ is a vector that measures the PSF anisotropy, and $P^{\rm sm}$ is
the smear polarisability tensor given in \citet{HFKS98}.
$p(\bm{\theta})$ 
can be estimated from images of stellar objects at position
$\bm{\theta}$ by noting that a 
star, denoted throughout this paper with $^*$, imaged
in the absence of PSF distortions has zero ellipticity:  
$\varepsilon^{*\, {\rm cor}}_{\alpha} = 0$. Hence,  
\be
p_{\mu} = \left(P^{\rm sm *} \right)_{\mu \alpha}^{-1}\,
\varepsilon^{*{\rm obs}}_{\alpha} .
\label{eqn:pmu}
\ee
The isotropic effect of the atmosphere and weight function can be accounted for
by applying the pre-seeing shear polarisability tensor correction $P^\gamma$,
as proposed by \citet{LK97}, such that
\be
\varepsilon^{\rm cor}_{\alpha} = \varepsilon^{s}_{\alpha} + P^\gamma_{\alpha
  \beta}\gamma_{\beta}\,,
\label{eqn:eseeing}
\ee
where $\varepsilon^{s}$ is the intrinsic source ellipticity and $\gamma$ is the
pre-seeing gravitational shear.
\citet{LK97} show that
\be
P^\gamma_{\alpha \beta} = P^{\rm sh}_{\alpha \beta} - P^{\rm
  sm}_{\alpha \mu} \left(P^{\rm sm *} \right)_{\mu \delta}^{-1} P^{\rm
  sh *}_{\delta \beta},
\label{eqn:Pgamma}
\ee
where $P^{\rm sh}$ is the shear polarisability tensor given in \citet{HFKS98}
and $P^{\rm sm *}$ and $P^{\rm sh *}$ are the stellar smear and shear
polarisability tensors respectively.  
Combining the PSF correction, equation (\ref{eqn:ecor}), and the $P^\gamma$
seeing correction, 
the final KSB+ shear estimator $\hat{\gamma}$ is given by
\be
\hat{\gamma}_{\alpha} = \left(P^\gamma \right)^{-1}_{\alpha
  \beta} \left[\varepsilon^{\rm obs}_{\beta} - P^{\rm
  sm}_{\beta\mu}p_{\mu}\right].
\label{eqn:shearest}
\ee
This method has been used by many of the authors although different
interpretations of the 
above formula have introduced some subtle differences
between each authors' KSB+ implementation.  For this reason we provide precise descriptions of
each KSB+ pipeline in the Appendix.

\subsection{K2K Method}
One drawback of the KSB+ method is that for non-Gaussian PSF distortions, the
KSB PSF correction is mathematically poorly defined.  \citet{K00} (K2K)
addresses this issue
by properly accounting for the effects of a realistic PSF.  It also
proposes measuring shapes from images that have been
convolved with a re-circularising PSF, where the re-circularising PSF is a
$90^\circ$ rotation of a modeled version of the PSF.
Section 2.3.6 of \citet{DahleK2K}
provides a condensed description of the K2K shear estimator which has been
applied to the STEP simulations by Dahle (HD).  

\subsection{Shapelets}
The shapelets formalism of \citet{shapelets1} allows galaxy images to be
decomposed into orthogonal basis functions which transform simply
under a variety of operations, in particular shear and
(de)convolution. The expansion is based on a circular Gaussian, but
inclusion of higher orders allows general shapes to be described well.

\citet{KK06} uses the shapelets formalism of \cite{shapelets1} 
to derive
individual shape estimators that differ from the
method of \citet{shapelets}.  We briefly review this method which is 
based on the `constant ellipticity object' estimator of \citet{Kuijken},
referring the reader to \citet{KK06} for further details.
Each galaxy image is fitted as an
intrinsically circular source that has been sheared and then smeared
by the PSF. These operations are efficiently expressed in terms of
shapelets as
\begin{equation}
G_{\rm model}= P\cdot(1+\gamma_1 S_1 + \gamma_2 S_2) \cdot C
\end{equation}
where $G_{\rm model}$ is the model for the galaxy image, $P$ is the
known PSF convolution operator (expressed as a matrix operating on
shapelet coefficients), $S_i$ are the first-order shear operators,
$\gamma_i$ are the shear distortions that are fitted, 
and $C$ is a general circular
source of arbitrary radial luminosity profile (expressed as a
superposition of shapelets).  
Note that $P$ is determined from stellar objects whose shapelet
coefficients are interpolated separately across the field of view to the
position of each observed galaxy.
Fitting this model to each observed
galaxy image yields a best-estimate $(\gamma_1,\gamma_2)$ 
shear distortion value for
each galaxy, which can then be averaged or correlated to yield shear
estimators. In this paper, we use 
$\gamma_i = \langle\gamma_i\rangle/(1-\langle\gamma^2\rangle)$ 
as an estimate for the shear from the ensemble population. 
The factor in the denominator is the response of the average ellipticity of 
a population of elliptical sources to an overall shear (BJ02).
To cope with possible centroiding errors, an arbitrary
translation is included in the fit as well. The uncertainties on the
pixel values of each galaxy image can be propagated into the shapelet
coefficients, and to the estimates of the $\gamma_i$.
This method is exact for galaxies that are intrinsically circular or
elliptical. \citet{Kuijken} shows that this method also works well
for galaxies whose ellipticity or position angle varies
with radius.

\subsection{Im2shape}
Im2shape (\citet{im2shape}, Bridle et al. 2005, in prep) 
fits a sum of elliptical Gaussians to each object image,
taking into account unknown background and noise levels.  This approach 
follows that suggested by \cite{Kuijken}.

{\it SExtractor} is used to define postage stamps containing each object\footnote{The postage stamps used for this analysis were $16 \times 16$ 
pixels centered
on the {\it SExtractor} position.} and
galaxies and stars are
selected from the size magnitude plot from the {\it SExtractor} output.
The galaxies are modeled by Im2shape 
using two concentric Gaussians, with 6 free parameters
for the first Gaussian, and 2 additional free parameters (size and
amplitude) for the second Gaussian. The noise is assumed to be uncorrelated,
Gaussian and at the same level for all 
pixels in the postage stamp. The background level is
assumed to be constant across the postage stamp.  
Including the noise and background
levels there are 10 free galaxy parameters in total. 
Two Gaussians are used for the stars in all the images, except for
PSF 2, for which the amplitude of the second Gaussian was found to be
so small that one Gaussian was used instead. Where two Gaussians were used
to fit the stars, the Gaussians were taken to have totally independent
parameters, with 12 free parameters for the Gaussians, plus the
noise and background levels, making 14 free parameters in total.
To estimate these free parameters fast and efficiently, Im2shape makes use of
the BayeSys engine (written by Skilling \& Gull).  This implements
Markov-Chain Monte Carlo sampling (MCMC) which 
is used to obtain samples from
the probability distribution of the unknown parameters.
Estimates of the free parameters are then taken from the mean value of
the parameter across the MCMC samples, and the uncertainties are taken
from the standard deviation.
With this data set the MCMC analysis takes $~\sim 15$ seconds per galaxy image
on the COSMOS\footnote{{\it www.damtp.cam.ac.uk.cosmos}, SGI Altix 3700, 1,3 GHz Madison processors} supercomputer.

To account for the PSF a grid
of $5 \times 5$ points was defined on each image, and the PSF at each point
was estimated by taking the median parameters of the nearest five stars (note
that Im2shape was run on all the stellar-like objects and cuts were then 
used to remove outliers).  For each galaxy,  the PSF shape was taken from the
grid point closest to the galaxy in question. The trial galaxy
parameters were then combined with the PSF parameters analytically to
calculate the convolved image shape. The intensity in the centre
of each pixel is calculated and this is corrected for the integration
over the pixel using the curvature of the Gaussian at the centre of
the pixel (for both star and galaxy shape estimation).
The final ellipticity values for each galaxy (equation~\ref{eqn:elliparam}) 
are found from averaging over all the MCMC samples. 
Only galaxies with ellipticity
uncertainties less than 0.25 were included in the final catalogue,
as for higher ellipticity uncertainties the error estimates are
less reliable resulting from the
probability distribution becoming
less Gaussian. To obtain an
estimate of the shear from these ellipticity estimates 
the ellipticities are weighted by the inverse
square of the ellipticity uncertainties added in quadrature with the
intrinsic ellipticity dispersion $\sigma_e$ of the galaxies, found to be
$\sigma_e = 0.2$. 

\subsection{Wittman method with {\it ellipto}}
\label{sec:witt}
This method uses a
re-circularising kernel to eliminate PSF anisotropy, and `adaptive'
moments (moments weighted by the best-fit elliptical Gaussian) to
characterise the ellipticity of the source galaxies.  It is a partial
implementation of BJ02, discussed in section~\ref{sec:BJ02}, and primarily
differs from BJ02 by using a simpler re-circularising kernel.    

{\it SExtractor} is used for initial object detection. {\it SExtractor} 
centroids
and moments are then input to the {\it ellipto program}
\citep{elipto,Smith_thesis}
which measures the adaptive moments.  {\it ellipto} also
re-measures the centroid and outputs an error flag when the centroid
differs from the SExtractor centroid. This typically happens with
blended objects or those with nearby neighbours, whose measured shapes
may not be trustworthy in any case.  Stars are selected with an
automatic routine which looks for a dense locus at a constant {\it ellipto}
size. The selection is then visually checked. In real data, $\sim$5\%
of images require manual tweaking of the star selection, although 
this manual stage was not required for the STEP simulations.  
The spatial variation of the adaptive moments is
then fit with a second-order polynomial for each CCD of each
exposure. This fit is then used to generate a spatially varying $3 \times 3$
pixel re-circularising kernel, following \citep{FischerTyson97}.  Note
that a $3 \times 3$ kernel may be too small to properly correct a well-sampled,
highly elliptical PSF; the practical limit appears to be $\sim$ 0.1
ellipticity.  In those cases, the re-circularisation step may be
applied iteratively, mimicking the effect of larger kernels.  For the
STEP simulations, only PSF 3 required a second iteration, but 
three iterations were applied to all PSFs.

After re-circularisation, the object detection and {\it ellipto}
measurements are repeated to generate the final catalogue.  Note that
object detection on the re-circularised image in principle eliminates
PSF-anisotropy-dependent selection bias.  Objects are rejected from
the final catalogue if: the {\it ellipto} error is non-zero; measured
(pre-dilution-correction) scalar ellipticity $> 0.6$ (simulations show
that, with ground-based seeing, most of these are blends of unrelated
objects); or size $< 120\%$ of the PSF size.  The adaptive moments are
then corrected for dilution by an isotropic PSF and a responsivity
correction using the formulae of BJ02.  Weighting is not applied to
the data. Note that this method
has been used for cluster analyses but not for any published cosmic
shear results.

\subsection{Bernstein and Jarvis Method: BJ02}
\label{sec:BJ02}
The Jarvis (MJ) and Nakajima (RN) methods each extend the
{\it ellipto} technique by methods detailed in BJ02. 
Both are based upon expansions of the galaxy and PSF shapes into a
series of orthogonal 2D Gaussian-based functions, the
Gauss-Laguerre expansion, also known as `polar
shapelets' in \citet{Masseyshapelets}.  Both the Jarvis (MJ) and 
Nakajima (RN) 
methods move beyond the approximation, inherent in both the
{\it ellipto} and KSB methods, that the PSF asymmetry can be
described as a first-order perturbation to a circular PSF.
The Jarvis (MJ)
method applies `rounding kernel' filters from size
$3\times3$ pixels and up to the images 
in order to null several asymmetric Gauss-Laguerre
coefficients of the PSF, not 
just the quadrupoles. Note that for PSF ellipticities 
of order $\sim 0.1$, a $3 \times 3$ pixel 
kernel is sufficient to round out stars up to approximately $30$ pixels in 
diameter.  The galaxy shapes are next measured by
the best-fit elliptical Gaussian; formulae proposed by \citet{Hirata},
are used to correct the observed shapes for
the circularising effect of the PSF.

The `deconvolution fitting method' by Nakajima (RN) 
implements nearly the full formalism proposed
by BJ02, which is further elaborated in Nakajima et al (2005, in
prep): the {\em intrinsic} shapes of galaxies are modeled 
as Gauss-Laguerre expansions (to $8^{\rm th}$ order).  These are then
convolved with the PSF and fit directly to the observed pixel
values in a similar fashion to \citet{Kuijken}. This should fully capture
the effect of highly 
asymmetric PSFs or galaxies, as well as the effects of finite
sampling.  Note that both methods use the weighting scheme described in
section 5 of BJ02.

A difference between the BJ02 approaches and the
\citet{shapelets} shapelets implementation is that the
latter uses a circular Gaussian basis set, whereas the BJ02
method shears the basis functions until they match the 
ellipticity of the galaxy.  This in principle eliminates the
need to calculate the `shear polarisabilities' that appear in
KSB.

\section{STEP Simulation Data}
\label{sec:data}
For this analysis we have created an artificial set of
survey images using the
{\it SkyMaker} programme\footnote{http://terapix.iap.fr/cplt/oldSite/soft/skymaker}.
A detailed description of this software and the galaxy catalogue
generator, {\it Stuff}\footnote{ftp://ftp.iap.fr/pub/from\_users/bertin/stuff}, 
can be found in \citet{erben} and Bertin \& Fouqu\'e (in 
prep) and we therefore only provide a brief
summary here.  In short, for a given cosmology and survey description,
galaxies are distributed in redshift space with a luminosity and
morphological-size  
distribution as defined by observational and semi-analytical relations.
Galaxies are made of a co-axial de Vaucouleurs-type spheroid bulge 
and a pure oblate circular exponential thin 
disk \citep[see][for details]{SExt}. 
The intrinsic flattening $q$ of spheroids is taken between $0.3$ and $1$,
and within this range follows a normal distribution with
$\langle q\rangle=0.65$ and $\sigma_q=0.18$ \citep{1970Sandage}. Note that 
we assume the same flattening 
distribution for bulges and ellipticals, even if there
is some controversy about this \citep{Boroson}.
Inclination angles $i$ are randomly assigned following a 
flat distribution, as expected from
uniformly random orientations with respect to the line of sight.
The apparent axis ratio $\beta$ 
is given by $\beta = \sqrt{q^2 \sin^2 i + \cos^2 i}$ for the spheroid
component,  
and given by $\beta = \cos i$ for the thin disk.
The bulge plus disk
galaxy is finally assigned a random position angle $\theta$ on the sky and 
the bulge and disk intrinsic ellipticity parameters are then calculated from
equation~\ref{eqn:elliparam}. 

It has been known for some time that pure oblate circular disks,
oriented with a flat distribution of
inclination angles, do not provide a
good match to the statistics from real disk galaxies
\citep{BinndeV,Grosbol,LML92}: in particular, 
observations show a striking deficiency of galaxies with zero
ellipticities. Although surface-brightness selection effects are not to
be ignored \citep[see for example][]{HuivA}, there is now general
agreement that this phenomenon mostly betrays intrinsic ellipticities of
disk planes. The origin of these intrinsic ellipticities is not
completely clear \citep[see][]{BinneyMerrifield}, and is thought to
originate partly from non-axisymmetric spiral structures and/or a 
tri-axial potential \citep{RixZar}. 
The simulations used in this analysis ignore these aspects, 
and the simulated galaxies are therefore intrinsically
`rounder' on average than real galaxies.  
This should not impact  
on the lensing analysis that follows, except in the cases where
weighting schemes are used 
that take advantage of the sensitivity of intrinsically
circular galaxies to measure weak lensing shear.  These schemes will 
have an apparent signal-to-noise advantage in the current simulations, which
is expected to decrease given real data.

\begin{table}
\begin{center}
\begin{tabular}{c|l|r}
PSF ID & PSF type & Ellipticity \\\hline\hline
0 & no anisotropy & $ 0.00$ \\\hline
1 & coma & $ \sim 0.04$\\\hline
2 & jitter, tracking error & $ \sim 0.08$\\\hline
3 & defocus & $ \sim 0.00 $\\\hline
4 & astigmatism &$ \sim 0.00 $\\\hline
5 & triangular (trefoil) & $ 0.00 $\\\hline\hline
\end{tabular}
\end{center}
\caption{The {\it SkyMaker} simulations are convolved with this series of uniform
  PSF models.}
\label{tab:psfs}
\end{table}

\begin{figure}
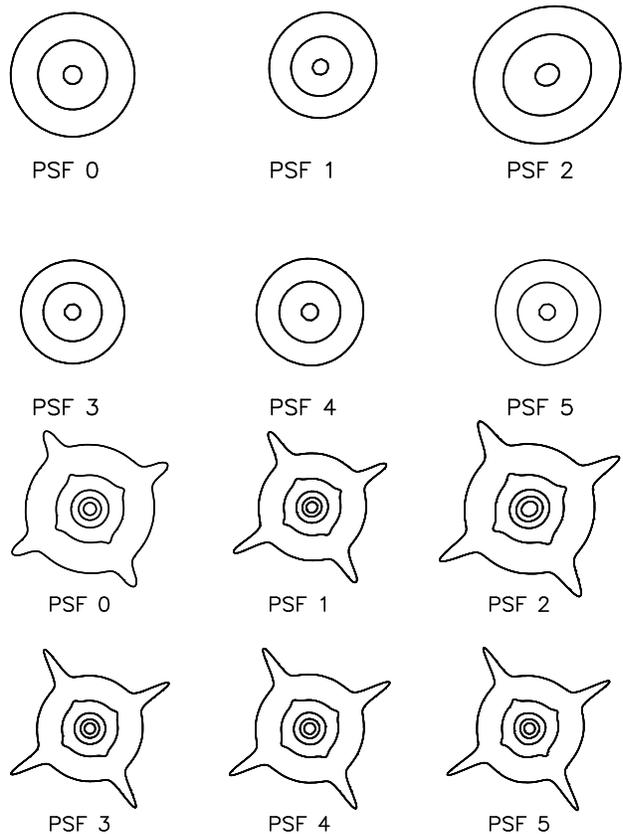

\begin{center}
\epsfig{file=cores.ps,width=5.5cm,angle=270,clip=}\\
\epsfig{file=outer.ps,width=5.5cm,angle=270,clip=}\\
\caption{{\it SkyMaker} PSF models, 
  as described in Table~\ref{tab:psfs}.  The upper
  panel shows the PSF core distortion, with contours marking $3\%$, $25\%$ and
  $90\%$ 
  of the peak intensity.  The lower panel shows the extended
  diffraction spikes,
  with contours marking $0.003\%$, $0.03\%$, $0.3\%$, $3\%$ and $25\%$ of the
  peak intensity.}
\label{fig:PSFS}
\end{center}
\end{figure}

A series of five different shears are applied to the galaxy catalogue by
modifying the observed intrinsic 
source ellipticity to create sheared galaxies where 
\be
e = \frac{e^{(s)} + g}{1 + g^* e^{(s)}}, 
\label{eqn:gstare}
\ee
\citep{SeitzSch97} and $g$ is the complex reduced shear.  For this set of
simulations, the 
convergence $\kappa = 0$, hence the reduced shear $g = \gamma/(1 - \kappa) =
\gamma$, where $\gamma_1 = (0.0, 0.005, 0.01, 0.05, 0.1)$, $\gamma_2=0.0$.
Sheared bulge and disk axial ratios and 
position angles are then calculated from
equation (\ref{eqn:elliparam}) and the model galaxy images are created. 
Stars are simulated assuming a constant slope of 0.3 per magnitude interval 
for the logarithm of differential stellar number counts down to and I-band
magnitude $I=25$.
Model galaxy images and stellar point sources are then convolved with a
series of six different optical PSFs that are listed in
Table~\ref{tab:psfs} and shown in Figure~\ref{fig:PSFS}.  These PSF models 
were chosen to provide a realistic representation of the
types of PSF distortions that are seen in ground-based observations,
through ray-tracing models of the optical plane.  They also 
include atmospheric
turbulence, where the seeing scale is chosen such that when the turbulence
is combined with the PSF anisotropy, all stars have FWHM of 0.9 arcsecs.
The ellipticity of the PSF
from real data is typically of the order of $5 \%$, which is 
similar to the coma model 
PSF 1.  PSF 2 which features a jitter or tracking error 
is very elliptical in comparison.  The other PSF
models test the impact of non-Gaussian PSF distortions. 
A uniform background with surface brightness 19.2 mag arcsec$^{-2}$ is added
to the image, chosen to match the I-band sky background at the
Canada-France-Hawaii Telescope site.
Poisson photon shot noise and Gaussian read-out noise is then applied.

The combination of $6$ different PSF types and $5$ different applied shears
gives $30$ different data sets where each set consists
of an ensemble of $64$ $4096 \times 4096$ pixel
images of pixel scale $0.206$ arcsecs.  For computational efficiency 
the data in each set stems from the same base catalogue, and as the sky noise
levels are the same for each data set, many of the parameters required for 
the {\it SExtractor} source detection software 
are the same for each data set.  
Aside from this time-saving
measure of setting some of the {\it SExtractor} 
source detection parameters only once, prior 
information about the simulations
have not been used in the cosmic shear analyses.   Each image contains 
$\sim 15$ galaxies per square arcminute resulting in low level 
shot noise from the
intrinsic ellipticity distribution at the $0.1\%$ level for each data set.
Stellar object density is $\sim 10$ stars per square
arcminute of which roughly 150 per image were sufficiently bright 
for the characterisation of the PSF.  This density of stellar objects is
slightly higher than that found with typical survey data and was chosen to 
aid PSF correction.  It does however increase the likelihood of stellar
contamination in the selected galaxy catalogue. Although the PSF is 
uniform across the field of view, uniformity has only been assumed in one case
(RN).  

The reader should note that the {\it SkyMaker} simulations should, in principle, 
provide an easy test of our methods 
as many shear measurement methods are based on the assumption
that the galaxy shape and PSF are smooth, elliptical 
and in some cases Gaussian.  In
reality the shapes of faint galaxies can be quite irregular and, particularly
in the case of space-based observations, the PSF can contain significant
structure.  In addition, the {\it SkyMaker} galaxies have reflection symmetry about
the centroid which could feasibly cause any symmetrical errors to vanish.
We should also note that some of the authors have previously 
used {\it SkyMaker} simulations to test their methods \citep[see][]{erben,HYGBHI}.
These issues will therefore be addressed by two future STEP publications
with the blind analysis of a more realistic set of artificial images
that use shapelet information to include complex galaxy morphology  
\citep{Massey_sim}.   With these shapelet simulations we will
investigate the shear recovery from 
ground-based observations (Massey et al. in prep) and
space-based observations (Rhodes et al. in prep).

\section{Analysis}
\label{sec:analysis}

\begin{table*}
\begin{center}
\begin{tabular}{c|c|c|c|c|c|c|c|c|c}
Author & Ngals (per ${\rm arcmin}^2$)&\% stars & \% false & \% stars$^\prime$ & \% false$^\prime$ & Software & SNR & S/$N_s$ & S/$N_s^\prime$\\ 
\hline\hline 
\input{Ngal_Nfalse.tab}
\end{tabular}
\end{center}
\caption{Table to compare the different number density of selected sources per
  square arcmin, Ngals, and the percentage of stellar contamination (\% stars)
  and false detections (\% false) in each authors'
  catalogue.  Each catalogue has been 
created using either the {\it SExtractor}
and/or the {\it hfindpeaks} software.  Where authors use object 
weights, the weighted percentage of stellar contamination (\% stars$^\prime$)
  and  
  false detections (\% false$^\prime$) are also listed. The final columns give
  estimates of the signal-to-noise of the resulting shear measurement as described in the text. SNR$=\gamma_i^{\rm true} / \sigma_\gamma$ is the signal-to-noise ratio of the shear measurement.  S/$N_s$ is the signal-to-shot-noise determined from the galaxies selected by each author. Where authors use object 
weights,  the signal-to-weighted-shot-noise S/$N_s^\prime$ is also determined. 
} 
\label{tab:Ngal_Nfalse}
\end{table*}

In this section we compare 
each authors' measured shear catalogues with the
input to each {\it SkyMaker} simulation.   
We match objects in each authors'
catalogue to the input galaxy and stellar catalogue, 
within a tolerance of 1 arcsec.  Table~\ref{tab:Ngal_Nfalse}
lists several general 
statistics calculated from the PSF model 0 (no anisotropy)
$\gamma = (0.005,0.0)$ set which is a 
good representation of the STEP
simulation data.  The source extraction method used by each author is
listed in Table~\ref{tab:Ngal_Nfalse} as well as 
the average number density of selected sources per square arcmin, Ngals. 
To minimise shot noise we wish to maximise the number of sources
without introducing false detections into the sample 
(note the percentage of false
detections listed in the `\% false' column 
in Table~\ref{tab:Ngal_Nfalse}) 
or contaminating the sample with stellar objects 
(note the percentage of stellar
contamination listed in the `\% stars' column in
Table~\ref{tab:Ngal_Nfalse}).  Both false 
objects and stars add noise which can dilute the average shear measurement.  
Typically the number of false detections are negligible and 
the stellar contamination is below 5\%.  The notable exception is
the Dahle (HD) method that suffers from strong stellar 
contamination for all PSF types, a problem that can easily be
improved upon in future analyses.
Where authors use object weights $w_i$ in their analysis, the weighted
percentage stellar
contamination (\% stars$^\prime= [\Sigma_{i =\rm stars} \, w_i \,/\, \Sigma_{i = \rm all} \,w_i] \times 100\%$) and weighted percentage of 
false object contamination (\% false$^\prime$) are also listed.
This shows, for example, that in the case of Hoekstra (HH), 
the $10\%$ stellar objects 
are given a very low weight and therefore do not significantly contribute to
the weighted average shear measurement.  

Average centroid offsets measured from each authors selected catalogues, 
were found to be $<0.001$ pixels for {\it SExtractor} based
catalogues and $\sim 0.005 \pm 0.001$ pixels for {\it hfindpeaks} based 
catalogues. Centroid accuracy is however 
likely to be data dependent, and S/N
dependent \citep[see][]{erben}.  Thus care should still
be taken in determining centroids to prevent the problems
described in \citet{vWb04} where 
errors in the {\it SExtractor} centroiding in one
field were found to be the source of strong B-modes on large scales.
Note that starting from version 2.4.3, {\it SExtractor} provides 
iterative, Gaussian-weighted centroid measurements {\tt XWIN\_IMAGE} 
and {\tt YWIN\_IMAGE} which have been shown to be even more accurate 
than previous {\it SExtractor} centroid measures (Bertin \& Fouqu\'e in 
prep).

For each data set we calculate the mean (weighted) shear  measured by
each author,  treating each of the $64$ images as an independent
pointing. We take the measured shear  for each data set $\gamma_i$ to
be the mean of the measurements from the $64$ images and assign an
error $\sigma_\gamma$ given by the error on the mean.  The final three
columns of Table~\ref{tab:Ngal_Nfalse} demonstrate the effect of
weights and galaxy selection on the signal-to-noise of the
measurement.  The signal-to-noise of the shear measurement is defined
as ${\rm SNR} = \gamma_i^{\rm true} / \sigma_\gamma$, where
$\gamma_i^{\rm true}$ is the input shear ($\gamma_1^{\rm true}= 0.005$
for the data analysed in Table~\ref{tab:Ngal_Nfalse}).  The
signal-to-shot-noise is defined as ${\rm S/N_s} = \gamma_i^{\rm true}
/ \sigma$ where $\sigma$ is the error on the mean galaxy
ellipticity $e$ (equation~\ref{eqn:elliparam}) measured from the $64$
images.  Note that the shot noise $\sigma$ is calculated from the known input
ellipticities of galaxies selected by each author.  The final column applies to
authors who use weights, where the signal-to-weighted-shot-noise is defined as
${\rm S/N_s^\prime} = \gamma_i^{\rm true} / \sigma^\prime$ where
$\sigma^\prime$ is the error on the mean weighted galaxy
ellipticity.

Several things can be noted from the signal-to-noise calculations.
Firstly, the high magnitude, as weak shear has not been
measured from data with SNR $> 10$.  One must not forget however that
if weak lensing shear was constant across large areas of sky, shear
would have been measured with such high signal-to-noise.  Secondly we
find that the signal-to-shot-noise ${\rm S/N_s}$ is not strongly
dependent on the number of galaxies used in the analysis.  We find
that instead the shot noise is more dependent on the galaxies that have
been selected in the analysis, but note that this statement is
unlikely to apply to data where the shear varies.   
Taking Im2shape (SB) and BJ02 (MJ) as
an example we find $\sim 2$ times as many galaxies selected for the
Im2shape (SB) analysis as for the BJ02 (MJ) analysis, but very similar
values for the signal-to-shot-noise ${\rm S/N_s}$.  As discussed in
section~\ref{sec:data} the distribution of galaxy ellipticities is
strongly non-Gaussian with more intrinsically round galaxies than is
seen in real data.  The galaxy selection of Im2shape (SB) results in a
smaller proportion of these intrinsically round galaxies being
included in the analysis increasing the $1\sigma$ variation of the
selected galaxy ellipticities.  Several of the KSB+ analyses make
galaxy selection based on galaxy ellipticity, removing the most
elliptical galaxies, again this reduces the shot noise, independent of the number of galaxies used in the analysis.  Lastly,
comparing the signal-to-shot-noise ${\rm S/N_s}$ and the
signal-to-weighted-shot-noise ${\rm S/N_s^\prime}$  we see the
effectiveness of some of the weighting schemes used in this analysis.
The BJ02 weighting scheme (MJ,RN) puts more weight on the
intrinsically round galaxies, this effective weighting scheme produces
the highest signal-to-noise measurements in the STEP analysis,
although see section~\ref{sec:disBJ02} for the implication of using this
aggressive weighting scheme.

\subsection{Calibration bias and PSF contamination}
\label{sec:calbias_PSF}
In this section we measure the levels of multiplicative 
calibration bias and additive 
PSF contamination in each authors' shear measurement.
Calibration bias will result from a poor correction for the
atmospheric seeing that circularises the images.  Selection bias and weight
bias are also forms of calibration bias which we investigate further in
sections~\ref{sec:selcbias} and~\ref{sec:weightbias}.  
PSF contamination will
result from a poor correction for the PSF distortion that coherently smears
the image.

\begin{center}
\begin{figure}
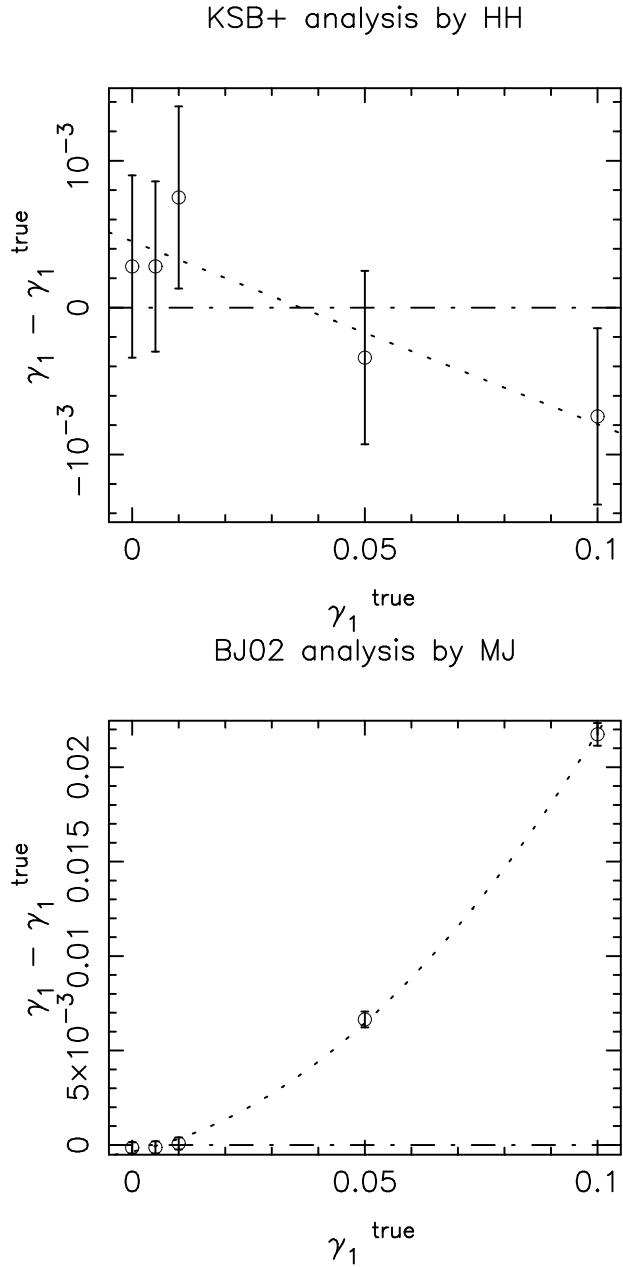

\epsfig{file=HH_example_plot.ps,width=8.2cm,angle=0,clip=}
\epsfig{file=MJ_example_plot.ps,width=8.2cm,angle=0,clip=}
\caption{Examples of two analyses of PSF 3 simulations using KSB+ (HH
implementation, upper panel) and BJ02 (MJ implementation, lower
panel) comparing the measured shear $\gamma_1$ and input shear
$\gamma_1^{\rm true}$.  The best-fit to equation~\ref{eqn:m1c1} is shown
dashed, and the optimal result (where $\gamma_1 = \gamma_1^{\rm
true}$) is shown dot-dashed.
Both analyses have additive errors that are consistent with shot noise
(fitted y-offset parameter $c$) and low ~1\% calibration errors (fitted
slope parameter $m$). The weighting scheme used in the BJ02 analysis
introduces a non-linear response to increasing input shear (fitted
quadratic parameter $q$), reducing the shear recovery accuracy for
increasing shear.  The accuracy of the KSB+ analysis responds linearly
to increasing input shear and so these results were re-fit with a
linear relationship, i.e. $q=0$.}
\label{fig:examples}
\end{figure}
\end{center}

We calculate the mean shear $\gamma_i$ for each data set 
as described above.  For each author and PSF type we then determine, from the
range of sheared images, the best-fit parameters to
\be
\gamma_1 - \gamma_1^{\rm true} = q(\gamma_1^{\rm true})^2 + m\gamma_1^{\rm
  true} + c_1 \,\,,
\label{eqn:m1c1}
\ee
where $\gamma_1^{\rm true}$ is the external shear applied to each image. Figure~\ref{fig:examples} shows fits to two example analyses of PSF 3 
simulations using KSB+ (HH implementation) and BJ02 (MJ implementation).  
In the absence of calibration bias we would expect $m=0$.  We would also
expect $c_1=0$ in
the absence of PSF systematics and shot noise, 
and $q=0$ for a linear response of the method
to shear.  In the case where the fitted parameter $q$ is consistent with zero,
we re-fit with a linear relationship, as demonstrated by the KSB+ example in 
figure~\ref{fig:examples}. 

For all simulations the external applied shear $\gamma_2^{\rm true} = 0$
and we therefore also measure for each PSF type 
$c_2 = \lag \gamma_2 \rag $, averaged over the range of sheared images.
In the absence of PSF 
systematics and shot noise, we would expect to find $c_2 = 0$.
From this analysis we found the values of $m$ and $q$ 
to be fairly stable to changes in PSF type and we therefore define a
measure of calibration bias to be $ \lag m \rag $ and a measure of non-linearity to be
$ \lag q \rag $ where the average is taken over the 6 different PSF sets.
We find the value of $ \lag c_i \rag $  averaged over the 6 different PSF sets
to be consistent with shot noise at the $0.1\%$ level for all authors, 
with the highest residuals seen with 
PSF model 1 (coma) and PSF model 2 (jitter).
We therefore 
define $\sigma_c$ as a measure of our ability to correct for all types of PSF
distortions, where $\sigma_c^2$ is the variance of $c_1$ and $c_2$ as measured
from the 6 different PSF models.  As the underlying galaxy distributions are
the same for each PSF this measure removes most of the 
contribution from shot noise, although the galaxy selection criteria
will result in slightly different noise
properties in the different PSF data sets.
$\sigma_c$ therefore provides a good estimate of the level of PSF
residuals in the whole STEP analysis.  A more complicated set of PSF distortions will be analysed in Massey et el (in prep) to address the issue of PSF-dependent bias more rigorously.

\begin{center}
\begin{figure}
\epsfig{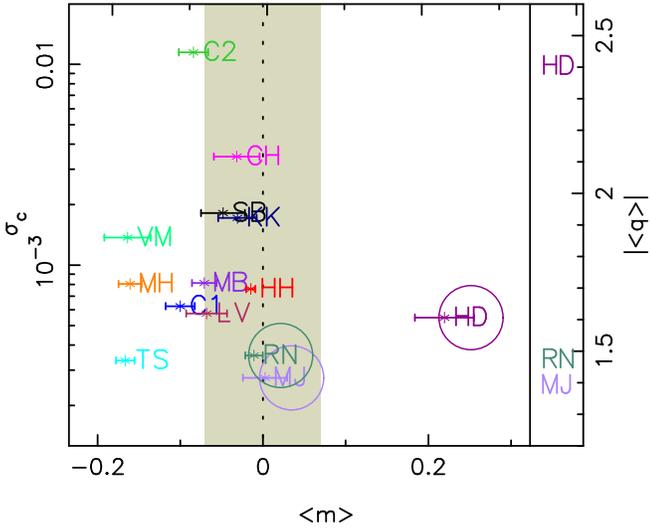}
\caption{Measures of calibration bias $\lag m \rag $, PSF residuals $\sigma_c$
  and 
  non-linearity $\lag q \rag $ for each author (key in
  Table~\ref{tab:methods}),  as described in the text.  
  For the non-linear cases where $\lag q \rag  \ne 0$ (points enclosed within a
  large circle), $\lag q \rag $ is
  shown with respect to the right-hand scale.
In short, the lower the value of $\sigma_c$, 
the more successful the PSF correction is at removing all types of PSF
distortion. The lower the absolute value of $\lag m \rag $, the lower the
  level of calibration bias. The higher the
$q$ value the poorer the response of the method to stronger shear.
  Note that for weak shear $\gamma < 0.01$, the impact of this quadratic term
  is negligible. 
Results in the shaded region suffer from less
than $7\%$ calibration bias. These results  
are tabulated in Table~\ref{tab:mcq}.}
\label{fig:mbar_sigc_qbar}
\end{figure}
\end{center}

Figure~\ref{fig:mbar_sigc_qbar} shows the measures of
PSF residuals $\sigma_c$ and calibration
bias $\lag m \rag $ for each author, where the author
key is listed in Table~\ref{tab:methods}.
For the non-linear cases where $q \ne 0$, denoted with
a circle, the best-fit $\lag q \rag $ parameter 
is shown with respect to the right-hand scale.
Results in the shaded region
suffer from less than $7\%$ calibration bias.  All methods which have been
used in a cosmological parameter cosmic shear analysis lie within this region.
With regard to PSF contamination, these results show that
PSF residuals are better than $1\%$ in all cases and are typically better than
$0.1\%$.  
Note that for clarity the results plotted in 
Figure~\ref{fig:mbar_sigc_qbar} are also tabulated in Table~\ref{tab:mcq}.

In the weak $\gamma \le 0.01$ regime,
the most successful method is found to be the BJ02 technique (MJ,RN)
producing percent level accuracy.  For stronger shear distortions, however, 
this methodology
breaks down which can be seen from the high $\lag q \rag $ value.  
This method is therefore 
unsuitable for low redshift cluster mass reconstructions where shear
distortions of $\sim 10\%$ are not uncommon, although see the discussion
in section~\ref{sec:disBJ02} for a solution to this issue of non-linearity. 
Over the full range of shear distortions tested, $0<\gamma<0.1$, 
the most successful method is found to be the 
Hoekstra implementation of the \citet{KSB} method (KSB+), 
producing results accurate to better than 2\%.  
All KSB+ pipelines are accurate to 
better than $\sim 15\%$ but the wide range of accuracy in these results that
are based on the same methodology is somewhat 
disconcerting.  It is believed that this spread results from the subtly
different interpretation and
implementation of the KSB+ method which we detail in the
Appendix. 
The results from the Dahle implementation of K2K (HD) are non-linear,
suffering from calibration bias at $\sim 20\%$ level for weak shear $\gamma <
0.01$.  
The Wittman/Margoniner method (VM) (see section~\ref{sec:witt}) fares as well
as the Hetterscheidt (MH) and Schrabback (TS) implementation of KSB+ with an
accuracy of $\sim 15\%$.
Im2shape \citep{im2shape} (SB) and the \citet{KK06} (KK) implementation of shapelets typically fare as well as the methods used in
cosmological parameter cosmic shear analyses
with an accuracy of $\sim 4\%$. 

\subsection{Selection Bias}
\label{sec:selcbias}
\begin{center}
\begin{figure}
\epsfig{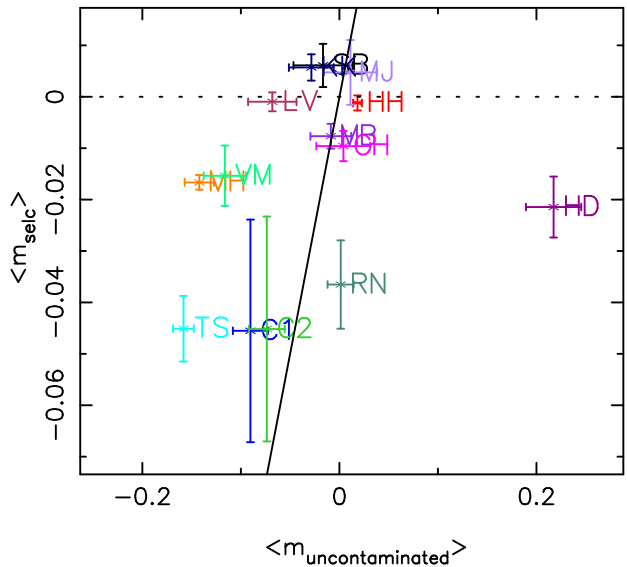}
\caption{Measures of selection bias $\lag m_{\rm selc} \rag$, for each author
  (key in Table~\ref{tab:methods}), as described in
  the text.  The lower the absolute value of $\lag m_{\rm selc}
  \rag$ the lower the level of selection bias.  Selection bias can be compared
  to 
the calibration bias $\lag m_{\rm uncontaminated} \rag$ 
measured from catalogues cleansed of false detections and
stellar contamination.    Unbiased shear measurement methods, 
where the shear is measured accurately but the source selection criteria are
  potentially biased, 
  would fall along the 1:1 line over-plotted.  These results  
are tabulated in Table~\ref{tab:mcq}.}
\label{fig:mselc_mbar}
\end{figure}
\end{center}

Selection bias is an issue that is potentially 
problematic for many different types of
survey analysis.  With weak lensing analyses,
which relies on the fact that when averaging over many
galaxies, the average source galaxy ellipticity $\lag e^{(s)} \rag = 0$,
removing even weak selection biases is particularly important.
When compiling source catalogues one should therefore consider any forms of
selection bias that may alter the mean ellipticity of the galaxy population.  This
bias could arise at the source extraction stage if there was a
preference to select galaxies oriented in
the same direction as the PSF \citep{K00} or galaxies that are
anti-correlated with the gravitational shear (and as a result appear more
circular) \citep{Hirata}.  Selection criteria applied after source
extraction could also bias the mean
ellipticity of the population if the selection has any dependence
on galaxy shape.  In this section we determine the level of selection bias by
measuring the unweighted mean intrinsic source
ellipticity $\lag e^{(s)} \rag$ (unlensed, equations~\ref{eqn:elliparam}
and~\ref{eqn:gstare}) from the 
`real' galaxies selected by each author for inclusion in their shear
catalogue (false detections are thus excised from the catalogue at this stage).
We follow a similar analysis to section~\ref{sec:calbias_PSF},
by determining for each author and each PSF type, from the range of sheared
images, the best-fit parameters to
\ba
\lag e_1^{(s)} \rag_{\rm selc} &=& m_{\rm selc} \gamma_1^{\rm true} + c_1^s \nn
\lag e_2^{(s)} \rag_{\rm selc} &=& c_2^s . 
\label{eqn:esm1c1}
\ea
$\lag m_{\rm selc} \rag$ averaged over the 6 different PSF data sets
gives a measure of the shear-dependent selection bias
and $(\sigma_c^s)^2$, the variance of $c_1^s$ and $c_2^s$ as measured
from the 6 different PSF models,  
gives a measure of the PSF-anisotropy-dependent selection bias.  We find
that PSF-anisotropy 
dependent selection bias is very low with $\sigma_c^s<0.001$ for
all methods.  Shear-dependent selection bias is $<1\%$ in most cases
with some notable exceptions in the cases of Clowe (C1 \& C2), 
Schrabback (TS), Dahle (HD) and Nakajima (RN) as shown on the vertical axis
of Figure~\ref{fig:mselc_mbar}.  The significant variation between the
different PSF data sets of $m_{\rm selc}$ measured with
the Clowe (C1 \& C2) catalogues 
suggests that the selection criteria of this method are affected by the PSF
type. 

Figure~\ref{fig:mselc_mbar} also shows the
value of $\lag m_{\rm uncontaminated}\rag$ determined from
equation (\ref{eqn:m1c1}) using the authors' measured 
shear catalogues now cleansed of false detections and stellar contamination,
with author-defined object weights.
With unbiased weights and an unbiased shear 
measurement method (where the shear is measured accurately but the source 
selection criteria are potentially biased),
points should fall along the 1:1 line
plotted.  We can therefore conclude from Figure~\ref{fig:mselc_mbar} that in
many cases the
calibration bias seen in section~\ref{sec:calbias_PSF} cannot be solely
attributed to selection bias.  See section~\ref{sec:diss} for a discussion on
sources of selection bias.
The results plotted in 
Figure~\ref{fig:mselc_mbar} are also tabulated in Table~\ref{tab:mcq}.
Comparing the calibration biases measured from the original catalogues $\lag m
\rag$ in Section~\ref{sec:calbias_PSF}, and from the `uncontaminated'
catalogues $\lag m_{\rm uncontaminated}\rag$ shows the impact of 
false detections and stellar contamination in each authors' catalogue.
Typically the impact is low with $< 3\%$ changes found for the average
measured shear of most authors.  One noticeable exception is the result from
the Brown (MB) pipeline, where the underestimation of the shear by $\sim 7\%$
is found to be predominantly 
caused by the diluting 
$\sim 7\%$ stellar contamination in the object catalogues. 

\subsection{Weight Bias}
\label{sec:weightbias}
In this section we investigate the impact of the different object-dependent 
weighting schemes
used by Bridle (SB), Clowe (C1 \& C2), Hetterscheidt (MH),
Hoekstra (HH), Kuijken (KK), Schrabback (TS) and Van Waerbeke (VW). 
All other methods use unit weights, except for the methods of 
Jarvis (MJ) and Nakajima (RN) which will be 
discussed at the end of this section.
An optimal weighting scheme should reduce the noise on a
measurement without biasing the results.   
Using the author defined weights we compare the average
unweighted and weighted mean intrinsic galaxy ellipticity, 
performing a similar analysis to sections~\ref{sec:calbias_PSF}
and~\ref{sec:selcbias}.   
For each author and PSF type we calculate from the range of sheared images,
the best fitting parameters to 
\be
\lag e_1^{(s)} \rag_{\rm selc} - \lag e_1^{(s)} \rag_{\rm selc}^{\prime}=
m_{\rm weight} \gamma_1^{\rm true} + c_1^w \,\,,
\label{eqn:ewm1c1}
\ee
where $\lag e_1^{(s)} \rag_{\rm selc}$ is an unweighted average and $\lag
e_1^{(s)} \rag_{\rm selc}^{\prime}$ is a weighted average.  
In the absence of PSF dependent weight bias, $c_1^w$ should be consistent with
zero and we find this to be the case for all the weighting schemes tested.
In the absence of
shear dependent weight bias, $m_{\rm weight}$ should be consistent with zero.
All weighting schemes are found to introduce low 
percent level bias as shown in Table~\ref{tab:mcq}, where $ \lag m_{\rm
  weight} \rag$ is averaged over the 6 different PSF models.  In most cases
these biases are small ($<2\%$) and we can therefore conclude the cases of
calibration bias seen in section~\ref{sec:calbias_PSF} cannot be solely
attributed to weight bias.  For percent level precision in future analyses
the issue of weight bias will need to be considered.  

The Jarvis (MJ) and Nakajima (RN) analyses make use of the 
ellipticity-dependent weighting
formulae in BJ02 Section 5.  This weighting scheme takes advantage of the
$e=0$ peak in the shape distribution of galaxies 
to improve the signal-to-noise of weak shear measurement.  
This is evidenced by the high signal-to-noise results with the 
Jarvis (MJ) and Nakajima (RN) methods 
as listed in Table~\ref{tab:Ngal_Nfalse}.  
Shearing the galaxies does change
the assigned weights, but the BJ02 formulae explicitly account for this
effect via a factor called the responsivity.  The non-linear response to
shear seen in the results of the Jarvis (MJ) and Nakajima (RN) 
methods is an undesirable consequence
of this weighting scheme which we discuss further in
section~\ref{sec:disBJ02}.  

\begin{table*}
\begin{center}
\begin{tabular}{c|r|c|c|r|r|r|c}
Author & $\lag m \rag$ & $\sigma_c$ & $\lag q \rag$ & $\lag m_{\rm
  uncontaminated} \rag$ 
& $\lag m_{\rm selc} \rag$ & $\lag m_{\rm weight} \rag$ &
$\sigma_8$ analysis ? \\\hline\hline
\input{m_bar_sigc_for_tab.dat}
\end{tabular}
\end{center}
\caption{Tabulated measures of calibration bias $\lag m \rag $, PSF residuals
  $\sigma_c$ 
  and non-linearity $\lag q \rag $ for each author (key in
  Table~\ref{tab:methods}), as described in Section~\ref{sec:calbias_PSF} and
  plotted in Figure~\ref{fig:mbar_sigc_qbar} .  
  For the non-linear cases where $\lag q \rag  \ne 0$, $\lag q \rag $ is
  listed.  `Uncontaminated' calibration bias $\lag m_{\rm
  uncontaminated} \rag $ is measured from object catalogues cleansed from
  stellar contamination and false object detections.  This can be compared to
  the measured selection bias $\lag m_{\rm selc} \rag $ as described in
  Section~\ref{sec:selcbias} and plotted in Figure~\ref{fig:mselc_mbar}.
  Weight bias $\lag m_{\rm weight} \rag $, described in
  Section~\ref{sec:weightbias}, is also tabulated.  For reference, the final
  column lists which pipelines have been used in cosmic shear analyses that
  have resulted in measurements of the amplitude of the matter power spectrum,
  $\sigma_8$, as detailed in Table~\ref{tab:sig8}.}
\label{tab:mcq}
\end{table*}

\subsection{Shear measurement dependence on galaxy properties}
\begin{center}
\begin{figure}
\epsfig{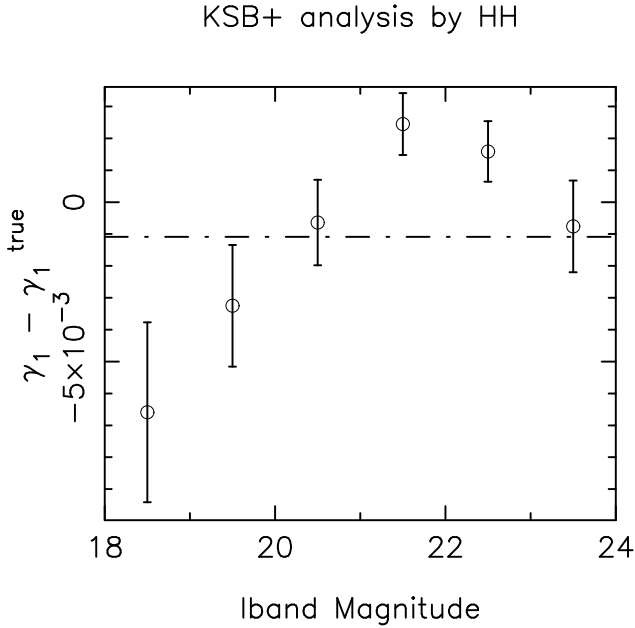}
\caption{An example plot of the difference between measured shear $\gamma_1$ and input shear $\gamma_1^{\rm true}$ as a function of galaxy $I$ band magnitude.  This plot is taken from the KSB+ analysis of HH using the PSF 0 simulations with an input shear $\gamma_1^{\rm true} = 0.05$.  The dot-dashed line shows the average $\gamma_1-\gamma_1^{\rm true}$ measured from the full galaxy sample.}
\label{fig:HH_mag}
\end{figure}
\end{center}

The simulations analysed in this paper were sheared uniformly across
the field-of-view.  In reality however,  the gravitational shear
experienced by each galaxy is dependent on position and more
importantly  redshift.  High redshift galaxies have a lower apparent
magnitude and smaller angular size when compared to their lower
redshift counterparts.  It is therefore important that shear
measurement methods are stable to changes in galaxy magnitude and
size.  For each author,  we measure the average shear as a function of
magnitude and input disk size.   In general,  we find that the average
shear binned as a function of magnitude and disk size varies $<1\%$ to
the average shear measured from the full data set, and an example plot
of shear measured as a function of galaxy magnitude is shown from the
KSB+ implementation of HH in Figure~\ref{fig:HH_mag}.  The dot-dashed
line shows the average $\gamma_1-\gamma_1^{\rm true}$ measured from
the full galaxy sample which is dominated by the faint magnitude
galaxies.  For this particular analysis the shear measured from bright
galaxies is slightly underestimated, and the shear from faint galaxies
is slightly overestimated.  The reader should note however that the
shear measured from each magnitude bin is $<1\sigma$ from the average
for all but one case and that for weaker input shears, this effect is even less
prominent.

Investigating the dependence of shear on galaxy properties we found that some 
methods introduced correlations between shear and magnitude, whilst others
between shear and disk size.  Interestingly however all methods revealed very
different dependencies on galaxy properties that we were unable to directly parameterise.  As such we cannot fully address the issue of shear measurement dependence on galaxy properties at this time.  For percent level precision in future analyses
this issue will certainly need to be revisited and it
will be addressed further in future STEP projects 
using simulations with constant shear and constant galaxy magnitude.

\section{Discussion}
\label{sec:diss}
In this section we discuss some of the lessons that we have learnt from the
first STEP initiative and highlight the areas where we can improve our
methods in future analyses.

\subsection{KSB+}
The subtle differences between the eight tested KSB+ pipelines, detailed in the
Appendix, introduces an interesting spread in the KSB+ results.  
Using the information in the Appendix, KSB+ users can now modify pipelines to 
improve their results.
The different ways of implementing KSB+ and the effect of using different
methods will 
be discussed in more detail in a future paper (Hetterscheidt et al in
prep), but comparing methods and results makes clear which interpretations of
the KSB+ method are best for ground-based data.  A good example of this is 
the PSF
correction method of Heymans (CH) and Clowe (C2) where the correction is
calculated as a function of galaxy size.  For ground-based data where the PSF
ellipticity is fairly constant at all isophotes (although note that this was
not the case with PSF 2), a PSF correction determined
only at the stellar size 
produces a less noisy and more successful PSF correction, as shown by the
success of the PSF correction by other KSB+ users.  This however would not 
necessarily be the case for space-based data where the PSF ellipticity varies
with size \citep[see for example][]{HymzGEMS} which will 
be tested in a future STEP analysis of simulated space-based observations.
The Schrabback (TS) method produces a more successful size-dependent PSF
correction by limiting the image region about stellar objects 
over which the PSF correction parameter
$p_\mu(r_g)$ is calculated
($\theta_{max}=3r_g^*$, see Appendix~\ref{app:Quad}).  
This measure reduces the noise on 
$p_\mu(r_g)$ thus improving the overall correction.

For several methods selection bias is well below the percent level from which
we can conclude that current source detection methods are suitable for weak
lensing analyses and that any selection bias seen with other methods 
has been introduced after the source extraction stage.  
The first clue to understanding the selection bias we see in some cases 
comes from 
comparing $\lag m_{\rm selc} \rag$ for the Hetterscheidt (MH) and Schrabback
(TS) results in Figure~\ref{fig:mselc_mbar}.  
These two analyses stem from the same {\it SExtractor} catalogue. 
The main differences between these two methods are the technique 
used to correct for the PSF distortion and the
catalogue selection criteria where Schrabback (TS) places more conservative
cuts on galaxy size defined by the ${\tt flux\_radius}$ parameter of {\it
  SExtractor}.  Whilst
there is no correlation within the simulations for intrinsic galaxy
ellipticity with disk size, we find that
the measured {\it hfindpeaks} $r_g$ parameter and the measured
${\it SExtractor}$  ${\tt flux\_radius}$ and ${\tt FWHM}$ parameters 
are somewhat 
correlated with galaxy ellipticity.  For this reason galaxy size 
selection criteria based
on  $r_g$, ${\tt flux\_radius}$ or ${\tt FWHM}$ 
will introduce a bias.   This finding is one of the
first lessons learnt from this STEP initiative which can now be improved upon
in future STEP analyses.

\subsection{K2K}
The Dahle (HD) K2K results appear noisier
than other pipelines which could result from an upper significance cut in 
order to remove big, bright galaxies, which in real data are at low redshift
unlensed galaxies.  This step rejects $\sim 24\%$ of the objects. 
The method is optimised for mosaic
CCD data with a high number of galaxies for each exposure, it therefore
suffers somewhat from the low number of objects in each $4096 \times 4096$
STEP  
image.  In addition, as a space-saving measure, images were stored in integer
format, this may have introduced some extra noise in the `re-circularised'
images.  In considering the success of K2K applied to STEP simulations one
should keep in mind that the man-hours invested in testing and 
fine-tuning KSB+ is at least an order of magnitude more than for any of the
other methods.  With the STEP simulations future tests and
optimisation are now feasible, the results of which will be demonstrated with
the next STEP analysis of shapelet based image simulations.

\subsection{Shapelets}
In the first, blind Kuijken (KK) analysis of the simulations all sources
were fitted to $8^{\rm th}$ order in shapelets, 
which gives a good fit to the
PSF-convolved sources. This, however, resulted in a systematic
underestimate of the shear amplitude of some 10\%. Later investigation
showed that even without any PSF smearing or noise, the ellipticity of
an exponential disk is only derived correctly if the expansion is
extended to $12^{th}$ order.   As this method has, to date, not been used 
in scientific analyses, it was decided that a re-analysis of the 
simulations with $12^{th}$ order shapelets would be permitted.
The results of the non-blind re-analysis are shown in
this paper. Using the higher order shapelet terms removed the
systematic underestimate for the high S/N sources. There is still a
tendency for noisy sources to have their ellipticities underestimated
however and this is still under investigation.

\subsection{Im2shape}
Im2shape uses MCMC sampling to fit elliptical Gaussians to the image.  Before
the STEP analysis it was believed that using too few
iterations in the MCMC analysis
would add noise to the ellipticities of each galaxy but would
not systematically bias them. It became apparent during this STEP
analysis 
however, that a bias is in fact introduced as the number of iterations is
decreased.  The number of iterations was chosen by systematically 
increasing the
number of iterations in the analysis of a subsample of the data until the
measured average shear converged.

\subsection{Wittman method with {\it ellipto}}

A post-STEP analysis of the shape catalogue revealed that the measured
galaxy shape distribution resulting from this method had rather
asymmetric tails.  The core of the distribution reflected the shear
much more accurately than did the mean of the entire distribution.
This method could thus be greatly improved by some type of weighting
or robust averaging scheme.  For example, a simple iterative 3$\sigma$
clip reduced the $15\%$ underestimate of the strongest applied shear, 
where $\gamma=0.1$, to an 8\%
underestimate, while rejecting only $2.2\%$ of the sources.  A slightly
harsher clip at $2.8 \sigma$ further reduced the underestimate to
$3.5\%$, while still rejecting only $3.9\%$ of the sources.  The stellar
contamination rate of $3.8\%$ is presumably responsible for the
remaining underestimate.  Note that the real data to which this method
has been applied is much deeper than the STEP simulations. The
stellar contamination rate would therefore be much lower, as the galaxy counts
rise more steeply with magnitude in comparison to the star counts.

Of course, one would prefer to understand the origin of the asymmetric
outliers rather than simply clipping them at the end.  A brief
analysis shows that they are not highly correlated with the obvious
variables such as photometric signal-to-noise or size relative to the
PSF.  Therefore a simple inverse-variance weighting scheme would not
be enough to solve the problem.  The prime task for improving this
method would thus be understanding the cause of this asymmetric tail
and developing a mitigation scheme.

\subsection{Bernstein \& Jarvis Method: BJ02}
\label{sec:disBJ02}
The ellipticity-dependent weighting scheme of BJ02 is responsible for the
significant increase in the signal-to-noise of the STEP shear measurements, as
shown in Table~\ref{tab:Ngal_Nfalse}.  It has, however, also been found to be
the cause of the non-linear response of the Jarvis (MJ) and Nakajima (RN)
methods to shear.
After the blind testing phase, the results of which are shown in this paper,
Jarvis (MJ) re-ran the analysis
with shape-independent weights finding a linear response to the range of weak
shears tested such that
the non-linearity parameter, $q$, measured by equation~\ref{eqn:m1c1} became
consistent with zero.  The signal-to-noise dropped, however, by a factor
of 1.5.   We can thus recommend that weak
shear studies use aggressive weights which help to probe small
departures of $\lag \gamma \rag$ from zero, while studies of stronger shear
regions use unweighted measurements to minimise the effects of non-linearity.

The false detections in the Nakajima (RN) analysis were investigated
and found to be either double objects detected by {\it SExtractor} as a single
object or
diffraction spikes.  Double object detections could be reduced by varying
SExtractor parameters to encourage the deblending of overlapping sources.  
When the data is taken in several exposures an additional measure to reduce
the number of false detections can be introduced.
This approach, taken by \citet{Jarvis}, demands that a source is
detected in at least two of the four exposures taken of each field.  The STEP
simulations were single exposure images and so this procedure could not be
implemented.  These false detections will generally be 
faint and highly elliptical in the case of diffraction spikes.  Thus, 
with the weighting scheme implemented in both the Jarvis (MJ)
and Nakajima (RN) analyses, these down-weighted objects do not 
affect the overall average measured shear.

\section{Conclusion}
\label{sec:conc}
In this paper we have presented the results of the 
first Shear TEsting Programme, where the accuracy of a wide range of
shear measurement methods were assessed. 
This paper has demonstrated that, for smooth galaxy light profiles, it is 
currently feasible to measure weak shear at percent level accuracy using the
\citet{Bernstein} method (BJ02) 
and the Hoekstra implementation of the KSB+ method.
It has
also shown how important it is to verify shear measurement software with
image simulations as subtle differences between each 
individuals implementation can result in discrepancy.  
We therefore strongly
urge all weak lensing researchers to subject their pipelines to a similar
analysis to ensure high accuracy and reliability in all future weak lensing
studies.  To this end the STEP simulations will be made available on request.

The removal of the additive 
PSF anisotropic distortion has been successful in all
methods, reduced to an equivalent shear of $\sim 0.001$ in most cases.   
Significant calibration bias is however 
seen in the results of some methods which can be explained only in part by
the use of biased weights and/or selection bias.   Using the simulations
analysed in this paper, errors can now be pin-pointed and corrected for, and
modifications will be introduced to remove sources of calibration error.
For authors using the KSB+ method, 
detailed descriptions have been given of each pipeline tested 
in this analysis to aid the improvement and development of future KSB+ methods.
One positive aspect of the KSB+ method is that its response to shear has
been shown to be very linear.  This is contrast to the BJ02 method
tested in this paper, where the ellipticity-dependent weighting scheme was
found to introduce a non-linear response to shear.
For this reason KSB+ or an unweighted version of the BJ02 method
is currently the preferred method for measuring weak shear around nearby
galaxy clusters.   Cosmic shear, on average, is very weak, but with the next
generation of cosmic shear surveys covering 
large areas on the sky and thus imaging regions of both high and low shear, 
cosmic shear measurement also 
requires a method that is linear in its response to shear.  Thus KSB+ or an
unweighted version of the BJ02 method is currently the preferred cosmic shear
measurement method.   In the weakest regime of galaxy-galaxy
lensing, the weighted
BJ02 method measures shear at a higher signal-to-noise with a better
accuracy than KSB+ and thus appears to be the most 
promising of the methods that have been tested in this analysis 
for galaxy-galaxy lensing studies.   

Selection bias has been shown to be consistent with zero in some cases, from
which we can conclude that current source detection methods are suitable for
weak lensing analyses.  Some object weighting schemes 
were found to be unbiased at the below percent level. 
The use of such schemes may however require
revision in the future when low level biases become important.  
All the methods tested were found to exhibit rather
different $<1\%$ dependences on galaxy magnitude and size.  
For real data where shear
scales with depth and hence magnitude and size, these issues will need to be
addressed. 

In this paper we
have provided a snapshot view of how accurately we can measure weak shear
today from galaxies with relatively simple galaxy morphologies.  
We are unable to answer the question, what method ought I to use to measure 
weak lensing shear?  KSB+, used with care, and BJ02 clearly fare well, 
but some of the methods tested here that are currently still in their 
development stage may still provide a better method in the future.
For the cosmic shear, galaxy-galaxy lensing 
and cluster-mass determinations published to
date, $\le7\%$ calibration errors are within
statistical errors and are certainly not dominant. 
$\sigma_c<0.01$ is also small enough to be sub-dominant in present work. 
We voice caution in explaining the $\sim 2\sigma$
differences in cosmological parameter estimation from cosmic shear studies
by the scatter in the results that we find in this
analysis.  The true reason is likely to be 
more complex involving source redshift uncertainties, residual systematics and 
sampling variance in addition to the calibration errors we have found.
Many of these sources of error will be significantly reduced with the next
generation of surveys where the large areas surveyed
will minimise sampling variance and the multi-colour data will provide a
photometric redshift estimate of the source redshift distribution.
The now widespread use of diagnostic
tools to determine levels of non-lensing residual 
distortions also allows for the 
quantification and reduction in systematic errors.  
Calibration errors, however, can only be directly 
detected through the analysis of image simulations.  

This first STEP analysis has quantified the current levels of calibration 
error, allowing for improvement in calibration accuracy in future 
shear measurement methods.  
The upcoming next generation of wide-field multi-colour optical surveys will
reduce statistical errors on various shear measurements to
the $\sim2\%$ level, requiring calibrations accurate to $\sim1\%$.
In the next decade, deep weak-lensing surveys of thousands of square
degrees will produce shear measurements that will be degraded by
calibration accuracies $\gtrsim 0.1\%$, well below even the precision
of the current STEP tests.  Similarly the additive errors represented
by $\sigma_c$ will ultimately have to be reduced to a level of
$\sigma_c < \approx 10^{-3.5}$ if this spurious
signal is to be below the measurement limits imposed by cosmic
variance of full-sky surveys.
The collective goal of the weak lensing community is now to 
meet these challenges.

The next STEP project will analyse a set of
ground and space-based image
simulations that include complex galaxy morphologies using a `shapelet'
composition \citep{Massey_sim}.  Initial tests with 
shapelet simulations suggest that complex
morphology rather complicates weak shear measurement for 
methods that assume Gaussian light profiles.  Further STEP projects
will address the issue of PSF interpolation and modeling, and the
impact of using different data reduction and 
processing techniques \citep{erben05}. These future STEP projects
will be as important as this first STEP analysis in order to gain
more understanding and further improve the accuracy of our methods.
We conclude with the hope that by using the shared technical
knowledge compiled by STEP, all future shear measurement 
methods will be able to reliably and accurately 
measure weak lensing shear. 

\section{Acknowledgments}
We thank TERAPIX (Traitement \'{E}l\'{e}mentaire, R\'{e}duction et Analyse des
PIXels de megacam) at the Institut d'Astrophysique de Paris for hosting the
{\it SkyMaker} simulations.  We also thank 
the Max-Planck-Institut f\"{u}r Astronomie for
financial and administrative support of STEP teleconferencing, and the Jet
Propulsion Laboratory for financial and administrative support of the STEP
workshop. CH is supported by a CITA National fellowship and
acknowledges financial support from GIF.  DB and MLB are supported by PPARC 
fellowships. SB used the UK National Cosmology Supercomputer Centre funded
by PPARC, HEFCE and Silicon Graphics / Cray Research.
HD is funded by a postdoctoral fellowship from the research
council of Norway.
TE acknowledges support from the German Science Foundation (DFG) 
under contract ER 327/2-1. KK acknowledges financial support 
provided through the European
Community's Human Potential Program under contract HPRN-CT-2002-00316, SISCO.
We thank Richard Ellis for helpful discussions about the STEP project and the referee for useful comments.

\bibliographystyle{mn2e}
\bibliography{ceh_2005}

\appendix
\section{KSB+ implementation}
The KSB+ method, used by a large percentage of the authors, has been shown in
this STEP analysis to produce
remarkably different results. In this Appendix, 
to aid the future understanding of these differences, we
detail how different authors have
implemented KSB+ with their weak lensing pipelines, as summarised in
Table~\ref{tab:KSB_details}.

\begin{table*}
\begin{center}
\input{KSB_table.tex}
\end{center}
\caption{The stages implemented 
by different authors using the KSB+ method
  described in section~\ref{sec:KSBmeth}.  Table notation; pix = pixel
  units; $P(r_g)$ implies
  that parameter $P$ is measured as a function of scale size 
  $r_g$; $P(r_g^*)$ implies
  that parameter $P$ is measured at the stellar scale size $r_g^*$. See the
  Appendix text for more details.} 
\label{tab:KSB_details}
\end{table*}

\subsection{Source detection, centroids and size definitions}
Most authors use the {\it SExtractor} software
\citep{SExt} to detect objects and define galaxy centroids.   
Exceptions are Hoekstra (HH) and Brown (MB) who use  {\it
    hfindpeaks} from the {\it imcat} software. 
The Gaussian weight scale length $r_g$ is then either set to the ${\tt
  flux\_radius}$ 
{\it SExtractor} parameter or the `optimal' $r_g$ value defined by {\it
  hfindpeaks}. 
Clowe (C1\&2) uses both pieces of software using a version of {\it hfindpeaks}
to determine the optimal weight scaling $r_g$ that keeps the centroid fixed to
the {\it SExtractor} co-ordinates.
Hetterscheidt (MH) and Schrabback (TS) measure half light radii $r_h$ and
refine the {\it SExtractor} centroids
using the iterative method described in \citet{erben}.

\subsection{Quadrupole moments and integrals}
\label{app:Quad}
The weighted ellipticity $\varepsilon$ (equation~\ref{eqn:ellipquad}), and
the smear and shear polarisability tensors
$P^{\rm sm}$ and $P^{\rm sh}$
are calculated for each object using software developed from the 
{\it imcat} subroutine {\it getshapes}.  The continuous integral formula are 
calculated from the discrete pixelised data by approximating the integrals as
discrete sums.   
The weighted ellipticity $\varepsilon$ is calculated from
the quadrupole moment which in its discrete form can be written as follows
\be
Q_{ij} = \frac{
\displaystyle\sum_{\theta_i, \theta_j = -\theta_{\rm max}}^{\theta_{\rm max}}
{\Delta\theta^2 \, W(\theta_i,\theta_j) \,
  I(\theta_i,\theta_j) \, \theta_i
  \theta_j}
} {
\displaystyle\sum_{\theta_i, \theta_j = -\theta_{\rm max}}^{\theta_{\rm max}}
{\Delta\theta^2 \, W(\theta_i,\theta_j) \,
  I(\theta_i,\theta_j)}},
\label{eqn:discquadmom}
\ee
where $\theta$ is measured, in pixel units, from the source centroid.
Table~\ref{tab:KSB_details} lists each authors' chosen values for $\theta_{\rm
  max}$ and $\Delta \theta$.  For real values of $\theta$, the intensity
$I(\theta_i,\theta_j)$, known at pixel positions, 
is estimated from a first-order interpolation over the four nearest pixels to
$(\theta_i,\theta_j)$ (denoted `interpolation' in
Table~\ref{tab:KSB_details}).  
The interpolation 
stage is by-passed by some authors by setting $\Delta \theta = 1$ pixel
and approximating $I(\theta_i,\theta_j)
\approx I({\rm Int}[\theta_i],{\rm Int}[\theta_j])$ (denoted `Approx' in
  Table~\ref{tab:KSB_details}), or by exchanging
the value of $\theta$, in the above formula, for its nearest integer value
${\rm Int}[\theta]$ 
(denoted `Integer' in Table~\ref{tab:KSB_details}). 
 $P^{\rm sm}$ and $P^{\rm sh}$ are functions of weighted moments, up to fourth
 order, that include $\theta_i\theta_j$ terms.  Some authors  
treat these second order terms in $\theta$ differently using the nearest 
integer values
of $\theta$ (denoted `Integer' in the $P^{\rm sh}$ and $P^{\rm sm}$ 
estimate column of 
Table~\ref{tab:KSB_details}).

\subsection{Anisotropic PSF modeling}
Stellar objects are selected by eye from the stellar locus in a
size-magnitude plane and are then used to produce a polynomial 
model of the PSF as a function of chip position.  
Hetterscheidt (MH), Heymans (CH) and Schrabback (TS) fit directly to $p_\mu$
(equation~\ref{eqn:pmu}) 
which, in the case of Heymans (CH) and Schrabback (TS), is measured
for varying $r_g$ \citep{HFKS98}.  This is 
in contrast to Hetterscheidt (MH) who
measures $p_\mu$ with $r_g = r_g^*$.  
Brown (MB), Clowe (C1\&2), Hoekstra (HH) and Van Waerbeke (LV) create models  
of $ \varepsilon^{*{\rm obs}}_{\alpha}$, $P^{\rm sm *}$ and $P^{\rm sh *}$
separately where for Brown (MB),  
and the first Clowe method (C1) stellar shapes are measured with $r_g = r_g^*$.
The second Clowe method (C2), the Hoekstra (HH) method 
and the Van Waerbeke (LV) method measures the stellar
parameters for varying $r_g$.  Note that the Van Waerbeke (LV) method fits
each component of the $P^{\rm sm *}$ and $P^{\rm sh *}$ tensors.
With PSF models in hand observed galaxy ellipticities are corrected according
to equation (\ref{eqn:ecor}).

\subsection{Isotropic $P^\gamma$ correction}

The application of the anisotropic PSF correction leaves 
an effectively isotropic distortion making objects rounder as a
result of both the PSF and the Gaussian weight function used to
measure the galaxy shapes. To correct for this effect and convert weighted
galaxy ellipticities $\varepsilon$ into unbiased shear estimators
$\hat{\gamma}$, we use the pre-seeing shear polarisability tensor $P^\gamma$,
equation (\ref{eqn:Pgamma}).  $P^\gamma$ is calculated for each galaxy from the
measured galaxy smear and shear polarisability tensors, $P^{\rm sm}$
and $P^{\rm sh}$, and a term that is dependent
on stellar smear and shear polarisability tensors; $\left(P^{\rm sm *}
\right)_{\mu \delta}^{-1} P^{\rm sh *}_{\delta \beta}$.  
Brown (MB) and the first method of Clowe (C1) use the stellar smear and shear
polarisability tensors measured with a Gaussian weight of scale size $r_g =
r_g^*$. 
Hetterscheidt (MH), Heymans (CH), Hoekstra (HH), 
Schrabback (TS), Van Waerbeke (LV) and the second method of Clowe (C2) 
calculate this stellar term  
as a function of smoothing scale $r_g$.  Comparing the C1 and C2 results
therefore 
demonstrates the impact of the inclusion of scale size at this stage.

$P^\gamma$ is a very noisy quantity, especially for small galaxies.  This
noise is reduced somewhat by treating $P^\gamma$ as 
a scalar equal to half its trace (note that the off diagonal terms of
$P^\gamma$ are typically an order of magnitude smaller than the diagonal
terms).  None of the  
methods tested in this analysis uses the full $P^\gamma$ tensor
correction (see \citet{erben} to compare the results achieved when
using a tensor and scalar $P^\gamma$ correction). 
In an effort to reduce the noise on $P^\gamma$ still
further, $P^\gamma$ is often fit as a function of $r_g$, although note that 
this fitting process has recently
been shown, with the Brown (MB) pipeline, 
to be dependent on which significance cuts are made when
selecting galaxies \citep{Massey}.
Table~\ref{tab:KSB_details} details which method is used by each author.  
In the case of
Clowe (C1\&2), $P^\gamma$ is also fit as a function of $\varepsilon$, and
with the method of Van Waerbeke, $P^\gamma$ is also fit as a function of
magnitude. 

In real data Hoekstra (HH) has previously found a clear dependence of 
$P^{\rm sh}$ on $\varepsilon$.
To correct for this shape dependence 
the Hoekstra pipeline multiplies $P^{\rm sh}$ 
by $(1 - \varepsilon^2/2)$ at the $P^\gamma$ correction stage. This
modification is not used in any of the other analyses.

\subsection{Weights}
\label{app:weights}
Some authors employ a weighting scheme in their analysis.  Hoekstra (HH) and
Van Waerbeke (LV) use
weights based on the error in the ellipticity measurement.  These weights are
derived in Appendix A1 of \citet{HFK00}. Clowe (C1\&2), Hetterscheidt (MH)
and   
Schrabback (TS) use a weighting
scheme based on the inverse of $\lag \gamma^2 \rag $ for all galaxies within a
given 
amount of $r_g$ and magnitude (TS,MH) or significance $\nu$ (C1\&2) of the
galaxy using a minimum of 20/50 (TS,MH/C1\&2) galaxies.   Note that 
this type of weighting applied to galaxies that have
experienced a constant shear will introduce a stronger bias that when the
same weights are applied to data where the shear varies.

\subsection{Selection criteria and calibration correction}
After applying the KSB+ method to the data each author has included a set of
selection criteria, listed in
Table~\ref{tab:KSB_details}.  These criteria are 
based on object significance $\nu$, 
`optimal' size $r_g$, half light radius $r_h$, 
observed ellipticity $\varepsilon^{\rm obs}$, corrected ellipticity
$\varepsilon^{\rm cor}$, measured shear $\gamma$, {\it SExtractor} stellar
class  ($1 = $ star, $0 = $ galaxy),  
measured/modeled $P^\gamma$ and so on.  The {\it imcat} software {\it
  getshapes} determines the offset of the flux averaged galaxy centroid (first
moment) from the given input galaxy centroid, 
scaled by the galaxy flux.  This measure, $d$, is
used by Clowe (C1\&2) to select `good' galaxies.  
A similar selection criterion is included in the methods
of Hetterscheidt (MH) and Schrabback (TS), where objects are only selected if
their iterative refinement of the centroid position 
converges and fixes the position to better than $2\times 10^{-3}$
pixels independently in $x$ and $y$. {\it imcat} also flags up saturated and
bad pixels 
which add noise to the quadrupole moments. Clowe (C1\&2) removes galaxies with
any saturated or bad pixels within $3r_g$ of the centroid.  

Brown (MB) includes a calibration
correction $\gamma_{\rm cor} = \gamma / 0.85$ as
suggested from the analysis of image simulations in \citet{Baconsims}.
Clowe (C1 \& C2) includes a close-pair calibration correction $\gamma_{\rm
  cor} = \gamma/0.95 $ to account for the diluting effect of blended objects. 
Normally Clowe visually inspects data to remove double objects
classified as a single source and sources with tidal tails in addition to
optical defects such as stellar spikes and satellite trails. This is feasible
with the typical amounts of data analysed in cluster lensing analyses.  For
wide-field cosmic shear surveys however
visual inspection becomes rather time consuming.  For this analysis 
Clowe therefore visually inspected 10 images from the simulation 
resulting in the rejection $\sim 5\%$ of the objects.
This process was found to increase 
the average shear measured in the visually inspected
images by $\sim 5\%$.  Thus Clowe includes a close-pair correction factor in
the STEP analysis to account for this effect in the whole simulation set.

\label{lastpage}

\end{document}

%% file: Ngal_Nfalse.tab
 SB & $ 18 $ & $    1.9 $ & $    3.8 $ & $    1.5 $ & $    3.9 $ & {\it SExtractor} & $  6$ & $  7$ & $  6$  \\ \hline
 MB & $ 14 $ & $    7.1 $ & $    0.1 $ & $    - $ & $    - $ & {\it hfindpeaks} & $  8$ & $ 10$ & $ -$  \\ \hline
 C1 & $ 12 $ & $    2.6 $ & $    0.0 $ & $    1.1 $ & $    0.0 $ & {\it hfindpeaks \& SExtractor} & $  9$ & $  9$ & $ 11$  \\ \hline
 C2 & $ 12 $ & $    2.7 $ & $    0.0 $ & $    1.2 $ & $    0.0 $ & {\it hfindpeaks \& SExtractor} & $  9$ & $  9$ & $ 11$  \\ \hline
 HD & $ 17 $ & $   44.8 $ & $  0.0 $ & $   - $ & $    - $ & {\it hfindpeaks} & $  7$ & $  8$ & $  -$  \\ \hline
 MH & $ 14 $ & $    3.9 $ & $    0.0 $ & $    2.4 $ & $    0.0 $ & {\it SExtractor} & $ 12$ & $ 11$ & $ 14$  \\ \hline
 CH & $ 12 $ & $    2.9 $ & $    0.0 $ & $    - $ & $    - $ & {\it SExtractor} & $  7$ & $ 11$ & $ - $  \\ \hline
 HH & $ 16 $ & $   10.8 $ & $    0.0 $ & $    0.1 $ & $    3.6 $ & {\it hfindpeaks} & $  8$ & $ 10$ & $ 11$  \\ \hline
 MJ & $  9 $ & $    0.0 $ & $    3.6 $ & $    0.0 $ & $    1.0 $ & {\it SExtractor} & $ 16$ & $  8$ & $ 22$  \\ \hline
 KK & $  9 $ & $    0.8 $ & $    0.0 $ & $    0.3 $ & $    0.0 $ & {\it SExtractor} & $ 10$ & $ 10$ & $ 12$  \\ \hline
 VM & $ 13 $ & $    3.8 $ & $    0.0 $ & $    - $ & $    - $ & {\it SExtractor} & $ 10$ & $ 10$ & $ -$  \\ \hline
 RN & $  9 $ & $    0.9 $ & $    0.4 $ & $    1.5 $ & $    0.1 $ & {\it SExtractor} & $ 19$ & $ 10$ & $ 24$  \\ \hline
 TS & $ 10 $ & $    1.4 $ & $    0.0 $ & $    0.9 $ & $    0.0 $ & {\it SExtractor} & $ 12$ & $ 11$ & $ 14$  \\ \hline
 LV & $ 13 $ & $    0.0 $ & $    0.0 $ & $    0.0 $ & $    0.0 $ & {\it SExtractor } & $ 11$ & $ 11$ & $ 12$  \\ \hline \hline

%% file: m_bar_sigc_for_tab.dat
 SB & $  -0.048 \pm  0.027$ & $  0.0018 $& - & $  -0.017 \pm  0.030 $ & $   0.006 \pm  0.004$ & $   0.007 \pm  0.002$ & $\times$ \\ \hline
 MB & $  -0.071 \pm  0.015$ & $  0.0008 $& - & $  -0.009 \pm  0.021 $ & $  -0.008 \pm  0.002$ &- & $\surd$\\ \hline
 C1 & $  -0.100 \pm  0.018$ & $  0.0006 $& - & $  -0.090 \pm  0.018 $ & $  -0.046 \pm  0.022$ & $   0.011 \pm  0.004$ & $\times$ \\ \hline
 C2 & $  -0.084 \pm  0.018$ & $  0.0115 $& - & $  -0.074 \pm  0.018 $ & $  -0.045 \pm  0.022$ & $   0.010 \pm  0.003$ & $\times$ \\ \hline
 HD & $   0.219 \pm  0.036$ & $  0.0005$ & $  -2.40 \pm   0.27 $ & $   0.217 \pm  0.028 $ & $  -0.021 \pm  0.006$ &- & $\times$ \\ \hline
 MH & $  -0.161 \pm  0.014$ & $  0.0008 $& - & $  -0.142 \pm  0.015 $ & $  -0.017 \pm  0.001$ & $   0.032 \pm  0.003$ & $\times$ \\ \hline
 CH & $  -0.032 \pm  0.028$ & $  0.0035 $& - & $   0.004 \pm  0.027 $ & $  -0.010 \pm  0.003$ &- & $\surd$\\ \hline
 HH & $  -0.015 \pm  0.006$ & $  0.0008 $& - & $   0.018 \pm  0.004 $ & $  -0.001 \pm  0.001$ & $   0.006 \pm  0.001$ & $\surd$\\ \hline
 MJ & $   0.002 \pm  0.027$ & $  0.0003$ & $   1.39 \pm   0.23 $ & $   0.011 \pm  0.027 $ & $   0.005 \pm  0.006$ &- & $\surd$\\ \hline
 KK & $  -0.031 \pm  0.023$ & $  0.0017 $& - & $  -0.029 \pm  0.023 $ & $   0.006 \pm  0.003$ & $   0.020 \pm  0.002$ & $\times$ \\ \hline
 VM & $  -0.164 \pm  0.028$ & $  0.0014 $& - & $  -0.116 \pm  0.021 $ & $  -0.015 \pm  0.006$ &- & $\times$ \\ \hline
 RN & $  -0.011 \pm  0.011$ & $  0.0004$ & $   1.47 \pm   0.09 $ & $   0.001 \pm  0.013 $ & $  -0.037 \pm  0.009$ &- & $\times$ \\ \hline
 TS & $  -0.167 \pm  0.011$ & $  0.0003 $& - & $  -0.158 \pm  0.010 $ & $  -0.045 \pm  0.006$ & $   0.024 \pm  0.003$ & $\times$ \\ \hline
 LV & $  -0.068 \pm  0.025$ & $  0.0006 $& - & $  -0.068 \pm  0.025 $ & $  -0.001 \pm  0.002$ & $   0.005 \pm  0.001$ & $\surd$\\ \hline

%% file: KSB_table.tex
\begin{tabular}{|l|c|c|c|c|}
\hline 
\hline 
KSB Author&
Brown &
Clowe &
Clowe &
Hetterscheidt \tabularnewline
\hline
\hline 
Key&
MB&
C1&
C2&
MH \tabularnewline
\hline 
Source Detection&
{\it hfindpeaks}&
{\it hfind} + {\it SExt} &
{\it hfind} + {\it SExt} &
{\it SExtractor}\tabularnewline
\hline 
PSF:&
$2^{\rm nd}$order&
$3^{\rm rd}$order &
$3^{\rm rd}$order &
$3^{\rm rd}$order\tabularnewline
2D polynomial&
to $\varepsilon^{*}$and&
to $\varepsilon^{*},P^{\rm sm*},P^{\rm sh*}$&
to $\varepsilon^{*},P^{\rm sm*},P^{\rm sh*}$&
to $p^{\mu}(r_{g}^*)$
\tabularnewline
model&
$P^{\rm sm*},P^{\rm sh*}$&
&
$(P^{\rm sh}/P^{\rm sm})(r_{g})$&
$3.5\sigma$ clipping
\tabularnewline
\hline 
Galaxy size $r_{g}$&
from {\it hfindpeaks}&
from {\it hfindpeaks}&
from {\it hfindpeaks}&
${\tt flux\_radius}$\tabularnewline
\hline 
Quadrupole estimate &
Approx&
Approx&
Approx&
Interpolation
\tabularnewline
$\theta_{\rm max}$ and $\Delta \theta$&
${\rm Int}[4 r_g]$, $1$ pix&
${\rm Int}[3 r_g]$, $1$ pix&
${\rm Int}[3 r_g]$, $1$ pix&
$3 r_g$, $0.25$ pix 
\tabularnewline
$P^{\rm sh}$ and $P^{\rm sm}$ estimate &
Approx&
Approx&
Approx& 
Interpolation
\tabularnewline
\hline 
$P^{\gamma}$ correction&
Fit of &
Fit $P_{ii}^{\gamma}(r_{g},e)$&
Fit $P_{ii}^{\gamma}(r_{g},e)$&
$\frac{1}{2}{\rm Tr}[P^{\gamma}]$\tabularnewline
&
$\frac{1}{2}{\rm Tr}[P^{\gamma}](r_{g})$&
$(P^{\rm sh}/P^{\rm sm})(r_{g}^{*})$&
$(P^{\rm sh}/P^{\rm sm})(r_{g})$ &
(no fit)\tabularnewline
\hline 
Weights&
none&
$<\gamma^{2}>^{-1}(r_{g},\nu)$ &
$<\gamma^{2}>^{-1}(r_{g},\nu)$ &
$<\gamma^{2}>^{-1}(r_{g},{\rm mag})$
\tabularnewline
\hline 
$\gamma$ correction &
Calibration &
Close-pair &
Close-pair &
\tabularnewline
 &
$\gamma_{\rm cor} = \gamma / 0.85$ & 
$\gamma_{\rm cor} = \gamma / 0.95$ &
$\gamma_{\rm cor} = \gamma / 0.95$ &
\tabularnewline
\hline
\hline 
Ellipticity cut&
$|\varepsilon_{obs}|$$\leq0.5$&
&
&
$|\varepsilon_{obs}|\leq0.8$
\tabularnewline
Size cut&
$r_{g}>r_{g}^{*}$&
$r_{g}^{*}<r_{g}<6$ pix&
$r_{g}^{*}<r_{g}<6$ pix&
$r_{h}>r_{h}^{*}$
\tabularnewline
Significance cut&
$\nu>5$&
$\nu>10$&
$\nu>10$&
\tabularnewline
$P^{\gamma}$ cut&
&
$P_{ii}^{\gamma}\geq0.15$&
$P_{ii}^{\gamma}\geq0.15$&
$\frac{1}{2}{\rm Tr}[P^{\gamma}]>0$
\tabularnewline
$\gamma$ cut&
&
&
&
\tabularnewline
Other&
&
$|d|<1$pix&
$|d|<1$pix&
$|d|<3$pix
\tabularnewline
&
&
SEx class $<$0.8&
SEx class $<$0.8&
\tabularnewline
&
&
No sat/bad pix&
No sat/bad pix&
\tabularnewline
\tabularnewline
\hline\hline 
KSB Author&
Heymans &
Hoekstra &
Schrabback &
Van Waerbeke \tabularnewline
\hline
\hline 
Key&
CH &
HH &
TS &
LV \tabularnewline
\hline 
Source Detection&
{\it SExtractor}&
{\it hfindpeaks}&
{\it SExtractor}&
{\it SExtractor}\tabularnewline
\hline 
PSF:&
$2^{\rm nd}$order&
$2^{\rm nd}$order&
$3^{\rm rd}$order&
$2^{\rm nd}$order \tabularnewline
2D polynomial&
to $p^{\mu}(r_{g})$ and &
to $\varepsilon^{*}(r_{g})$, &
to $p^{\mu}(r_{g})$&
to $\varepsilon^{*}(r_{g})$
\tabularnewline
model&
$(P^{\rm sm*})_{\alpha\beta}^{-1}P_{\beta\gamma}^{sh*}(r_{g})$&
$P^{\rm sm*}(r_{g})$ and $P^{\rm sh*}(r_{g})$&
&
$P^{\rm sm*}(r_{g})$ and $P^{\rm sh*}(r_{g})$
\tabularnewline
\hline 
Galaxy size $r_{g}$&
${\tt flux\_radius}$&
from {\it hfindpeaks}&
${\tt flux\_radius}$&
${\tt flux\_radius}$\tabularnewline
\hline 
Quadrupole estimate &
Approx&
Interpolation&
Interpolation&
Approx
\tabularnewline
$\theta_{\rm max}$ and $\Delta \theta$&
${\rm Int}[4 r_g]$, $1$ pix&
&
$3 r_g$, $0.25$ pix&
${\rm Int}[4 r_g]$, $1$ pix
\tabularnewline
$P^{\rm sh}$ and $P^{\rm sm}$ estimate &
Integer&
Interpolation&
Interpolation&
Approx
\tabularnewline
\hline 
$P^{\gamma}$ correction&
$\frac{1}{2}{\rm Tr}[P^{\gamma}]$&
$P^{\rm sh} \rightarrow (1-\varepsilon^2/2)P^{\rm sh}$ &
$\frac{1}{2}{\rm Tr}[P^{\gamma}]$&
Fit in $(r_{g},{\rm mag})$\tabularnewline
&
(no fit)&
Fit to &
(no fit)&
to $\frac{1}{2}{\rm Tr}[P^{\gamma}]$\tabularnewline
&
&
$\frac{1}{2}{\rm Tr}[P^{\gamma}](r_{g})$&\tabularnewline
\hline 
Weights&
none&
Hoekstra et al.&
$<\gamma^{2}>^{-1}(r_{g},{\rm mag})$&
Hoekstra et al.\tabularnewline
&
&
eqn A8,9&
&
eqn A8,9\tabularnewline
\hline
$\gamma$ correction &
\tabularnewline
\tabularnewline
\hline
\hline 
Ellipticity cut&
$|\varepsilon_{obs}|\leq0.5$&
&
$|\varepsilon_{cor}|\leq0.8$&
\tabularnewline
Size cut&
$1.2r_{g}^{*}<r_{g}<7$pix&
$r_h$ selection&
$r_{h}>1.2r_{h}^{*}$&
\tabularnewline
Significance cut&
$\nu>10$&
$\nu>5$&
&
$\nu>15$\tabularnewline
$P^{\gamma}$ cut&
&
&
$\frac{1}{2}{\rm Tr}[P^{\gamma}]>0$&
\tabularnewline
$\gamma$ cut&
$|\gamma|<2$&
&
&
\tabularnewline
Other&
Close pairs&
&
$|d|<3$ pix&
\tabularnewline
&
$<10$pix &
&
&
\tabularnewline
&
removed&
&
\tabularnewline
\hline
\end{tabular}